\documentclass[12pt]{article}

\usepackage{latexsym}
\makeatletter
\def\newleaf{\newpage
\newcount\tmp
\tmp=\c@page
\divide\tmp by 2
\multiply\tmp by 2
\ifnum\c@page=\tmp
~\newpage
\fi
}
\makeatother

\expandafter\ifx\csname proofnewpage\endcsname\relax
\let\proofnewpage=\relax
\fi

\def\color[#1]#2{}

\long\def\nop#1{}

\def\comment{\edef\cps{\the\parskip} \parskip=0.5cm \begingroup \tt}
\def\endcomment{\endgroup \vskip 0.5cm \parskip=\cps}
\def\separator{\vskip 1cm\hrule\vskip 1cm}

\expandafter\ifx\csname shortcite\endcsname\relax

\fi

\newbox\current

\long\def\plframebox#1{
\setbox\current\vbox{#1}		% set box

\vbox to \ht\current {\hrule\vss
\hbox to \wd\current {%
\vrule \hss\box\current\hss \vrule}
\vss\hrule }
}

\long\def\eatpar#1{%
\ifx#1\par                      % se il token e' \par
\let\nextmove=\eatpar           % rimetti \eatpar in coda
\else
\let\nextmove=#1%               altrimenti, rimetti il token in coda
\fi
\noexpand\nextmove%             il token o \eatpar viene rimesso in coda
}

\def\modifymargins#1#2{
\newdimen\addtoh
\newdimen\addtow
\addtoh=#1
\addtow=#2

\advance\topmargin by -\addtoh
\multiply\addtoh by 2
\advance\textheight by \addtoh

\advance\oddsidemargin by -\addtow
\advance\evensidemargin by -\addtow
\multiply\addtow by 2
\advance\textwidth by \addtow
}

\begingroup
\catcode`\~=11
\gdef\centertilde#1{\lower #1pt\hbox{~}}
\endgroup

\newcount\currenttime
\newcount\hour
\newcount\minute

\def\printtime{%
\currenttime=\time
\hour=\currenttime
\divide\hour by 60
\minute=-\hour
\multiply\minute by 60
\advance\minute by \currenttime
\the\hour:\ifnum\minute<10 0\fi\the\minute
}

\begingroup
\makeatletter
\global\let\@@date=\@date
\gdef\@date{\@@date\ --- \printtime}
\endgroup

\def\oggi{\number\day\space 
\ifcase\month\or
Gennaio\or Febbraio\or Marzo\or Aprile\or Maggio\or Giugno\or
Luglio\or Agosto\or Settembre\or Ottobre\or Novembre\or Dicembre\fi
\space \number\year}

\newcounter{rmexample}

\def\proof{\noindent {\sl Proof.\ \ }}

\def\qed{\hfill{\boxit{}}
  \ifdim\lastskip<\medskipamount \removelastskip\penalty55\medskip\fi}
\def\qedn#1{\hfill{\boxit{}$_#1$}
  \ifdim\lastskip<\medskipamount \removelastskip\penalty55\medskip\fi}
\long\def\boxit#1{\vbox{\hrule\hbox{\vrule\kern3pt
                  \vbox{\kern3pt#1\kern3pt}\kern3pt\vrule}\hrule}}

\def\ie{i.e.}
\def\eg{e.g.}
\def\wrt{w.r.t.}

\def\l{\langle}
\def\r{\rangle}

\def\mod{M\!od}

\def\var{V\!ar}

\def\true{{\sf true}}
\def\false{{\sf false}}

\def\S#1{\mbox{$\Sigma^p_{#1}$}}
\def\P#1{\mbox{$\Pi^p_{#1}$}}

\def\Dlog#1{\mbox{$\Delta^p_{#1}[\log n]$}}

\def\profont{\sf}

\def\x3c{{\profont x3c}}

\def\possnewtheorem#1#2{
\expandafter\ifx\csname #1\endcsname\relax
\newtheorem{#1}{#2}
\fi
}

\def\possnewtheoremthree#1[#2]#3{
\expandafter\ifx\csname #1\endcsname\relax
\newtheorem{#1}[#2]{#3}
\fi
}

\possnewtheorem{theorem}{Theorem}
\possnewtheorem{corollary}{Corollary}
\possnewtheorem{lemma}{Lemma}
\possnewtheoremthree{proposition}[theorem]{Proposition}
\possnewtheorem{definition}{Definition}
\possnewtheorem{question}{Question}
\possnewtheorem{example}{Example}
\possnewtheorem{nontheorem}{Counterexample}
\possnewtheorem{property}{Property}
\possnewtheorem{assumption}{Assumption}
\possnewtheorem{conjecture}{Conjecture}
\possnewtheorem{notation}{Notation}
\newenvironment{theorem*}[1]{{\noindent \bf Theorem~#1}\begin{it}}{\end{it}\

}

\def\circ{C\!I\!RC}

\ifx \isproof x
\else
\long\def\comment#1\endcomment{}
\fi

\def\cnf{CNF}
\def\twocnf{2CNF}

\modifymargins{40pt}{40pt}

\title{Redundancy in Logic III: Non-Mononotonic Reasoning}
\author{Paolo Liberatore%
\thanks{Universit\`a di Roma ``La Sapienza''.
Email: {\tt paolo@liberatore.org}}}

\begin{document}

\maketitle

%%%%%%%%%%%%% begin:abstract %%%%%%%%%%%%%%
\begin{abstract}

Results about the redundancy of circumscriptive and default
theories are presented. In particular, the complexity of
establishing whether a given theory is redundant is
establihsed.

\end{abstract}
%%%%%%%%%%%%% end:abstract %%%%%%%%%%%%%%
 %

\ifx \isproof x
\let\sectionnewpage=\relax
\tableofcontents
\let\sectionnewpage=\newpage
\let\subsectionnewpage=\newpage
\fi

%%%%%%%%%%%%% begin:introduction %%%%%%%%%%%%%%
\section{Introduction}

In this paper, we study the problem of whether a
circumscriptive \cite{mcca-80} or default \cite{reit-80}
theory is redundant, that is, it contains unnecessary parts.
Formally, a theory is redundant if it is equivalent to one
of its proper subsets; a part is redundant in a theory if
the theory is not semantically changed by the removal of the
part. The redundancy of propositional theories in \cnf,
\twocnf, and Horn form has already been analyzed in other
papers \cite{libe-05,libe-redu-2} where motivations are also
given. Other problems related to redundancy have been
considered by various authors
\cite{gins-88-a,schm-snyd-97,meye-stoc-72,maie-80,ausi-etal-86,hamm-koga-93,hema-wech-97,lang-marq-00,uman-97,gott-ferm-93,buni-zhao-05,papa-wolf-88,flei-etal-02,brun-03}.

Circumscription and default logic are two forms of
non-monotonic reasoning, as opposite to classical logic,
which is monotonic. A logic is monotonic if the consequences
of a set of formulae monotonically non-decrease with the
set. In other words, all formulae that are entailed by a set
are also entailed by every superset of it. Circumscription
and default logic do not have this property, and are
therefore non-monotonic.

The difference between monotonic and non-monotonic logic is
important in the study of redundancy. In propositional
logic, as in all forms of monotonic logic, if a set does not
entail a formula, the same is true for all of its subsets.
As a result, if $\Pi$ is a set of clauses, $\gamma$ is one
of its clauses, and $\Pi \backslash \{\gamma\}$ is not
equivalent to $\Pi$, no subset of $\Pi \backslash
\{\gamma\}$ is equivalent to $\Pi$. In the other way around,
if all clauses of a set of clauses $\Pi$ are irredundant,
then $\Pi$ is irredundant. We call this property {\em local
redundancy}. The converse of this property is obviously true:
if a clause is redundant in a formula, the formula is
redundant because it is equivalent to the subset composed of
all its clauses but the redundant one.

Local redundancy holds for all monotonic logic. In
nonmonotonic logics, removing a clause from a formula might
result in a decrease of the set of consequences, which can
however grow to the original one when another clause is
further removed. However, some nonmonotonic logics have the
local redundancy property. We prove that local redundancy
holds for circumscriptive entailment and for the redundancy
of the background theory in default logic when all defaults
are categorical (prerequisite-free) and normal. In the
general case, default logic does not have the local
redundancy property.

Since redundancy is defined in terms of equivalence (namely,
equivalence of a formula with a proper subset of it), it is
affected by the kind of equivalence used. In particular,
equivalence can be defined in two ways for default logic:
equality of extensions and equality of consequences. This
lead to two different definitions of redundancy in default
logic.

Regarding the complexity results, we show that checking
whether a clause is redundant in a formula according to
circumscriptive inference is \P{2}-complete. For default
logic, we mainly considered redundancy in the background
theory according to Reiter semantic using both kinds of
equivalence, but we also considered justified default logic
\cite{luka-88}, constrained default logic
\cite{scha-92,delg-scha-jack-94}, and rational default logic
\cite{miki-trus-95}. The results are as follows: the
redundancy of a clause in the background theory is
\P{2}-complete and \P{3}-complete for equivalence based on
extensions and consequences, respectively. The problems of
redundancy of the background theory are \S{3}-complete and
\S{4}-complete, respectively. The proofs of the latter two
results are of some interest, as they are done by first
showing that the problems are \P{2}-complete and
\P{3}-complete, respectively, and then showing that such
complexity results can be raised of one level in the
polynomial hierarchy. This technique allows for a proof of
hardness for a class such as \S{4}\  without involving
complicated QBFs such as $\exists W \forall X \exists Y
\forall Z.F$.

We also considered the redundancy of defaults in a default
theory. We show that these problems are at least as hard as
the corresponding problems for the redundancy of the
background theory for Reiter and justified default logics.

%%%%%%%%%%%%% end:introduction %%%%%%%%%%%%%%
 %

%%%%%%%%%%%%% begin:preliminaries %%%%%%%%%%%%%%
\section{Preliminaries}

If $\Pi$ and $\Gamma$ are sets, $\Pi \backslash \Gamma$
denotes the set of elements that are in $\Pi$ but not in
$\Gamma$. This operator is often called set subtraction,
because the elements of $\Gamma$ are ``subtracted'' from
$\Pi$. An alternative definition of this operator is: $\Pi
\backslash \Gamma = \Pi \cap \overline \Gamma$, where
$\overline \Gamma$ is the complement of $\Gamma$.

All formulae considered in this paper are propositional and
finite Boolean formulae over a finite alphabet. We typically
use formulae in \cnf, that is, sets of clauses. We simply
refer to sets of clauses as formulae. We assume that no
clause is tautological (\eg, $x \vee \neg x$): formulae
containing tautological clauses can be simplified in linear
time. By $\var(\Pi)$ we mean the set of variables mentioned
in the formula $\Pi$.

In some places, we use the notation $\neg \gamma$, where
$\gamma$ is a clause, to denote the formula $\{\neg l ~|~ l
\in \gamma\}$. Note that $\gamma$ is a clause, while both
$\{\gamma\}$ and $\neg \gamma$ are formulae (sets of
clauses). A clause is positive if and only if it contains
only positive literals.

A propositional model is an assignment from a set of
propositional variables to the set $\{\true,\false\}$. We
denote a model by the set of variables it assigns to
$\true$. We use the notation $\mod(\Pi)$ to denote the set
of models of a formula $\Pi$. We sometimes use models as
formulae, \eg, $\Pi \wedge \omega$ where $\Pi$ is a formula
and $\omega$ is a model. In the context where a formula is
expected, a model $\omega$ represents the formula $\{x ~|~ x
\in \omega\} \cup \{\neg x ~|~ x \not\in \omega\}$. If $\Pi$
is a formula and $\omega_X$ is a model over the variables
$X$, we denote by $\Pi|_{\omega_X}$ the formula obtained by
replacing each variable of $X$ with its value assigned by
$\omega_X$ in $\Pi$.

A quasi-order is a reflexive and transitive relation
(formally, a quasi-order is a pair composed of a set and a
reflexive and transitive relation on this set, but the set
will be implicit in this paper). The set containment
relation $\subseteq$ among models is a quasi-order.
According to our definition, a model is a set of positive
literals; as a result $M \subseteq M'$ holds if and only if
$M$ assigns to false all variables that $M'$ assign to
false.

A clause is (classically) redundant in a \cnf\  formula
$\Pi$ if $\Pi \backslash \{\gamma\} \models \gamma$. A \cnf\
formula is (classically) redundant if it is equivalent to
one of its proper subsets. Propositional logic has the local
redundancy property: a formula is redundant if and only if
it contains a redundant clause. The local redundancy
property is defined as follows.

\begin{definition}[Local redundancy]

A logic has the local redundancy property if, in this logic,
a theory is redundant only if it contains a redundant
clause.

\end{definition}

Propositional logic has the local redundancy property. This
is however not true for all logics.

%%%%%%%%%%%%% end:preliminaries %%%%%%%%%%%%%%
 %

%%%%%%%%%%%%% begin:circ %%%%%%%%%%%%%%
\section{Circumscription}

\def\x{ispell ignore this part \[}
\def\circ{{\rm CIRC}}
\def\infcirc{\models_M}
\def\x{\]}

Circumscriptive inference is based on the minimal models of
a theory, \ie, the models that assign the maximum quantity
of literals to false. Formally, we define the set of minimal
models as follows.

\begin{definition}

The set of minimal models of a propositional formula $\Pi$,
denoted by $\circ(\Pi)$, is defined as follows.

\[
\circ(\Pi)
=
\min_\subseteq(\mod(\Pi))
\]

\end{definition}

We define $\circ(\Pi)$ to be a set of models instead of a
formula, although the latter is more common in the
literature. Circumscriptive entailment is defined like
classical entailment but only minimal models are taken into
account.

\begin{definition}

The circumscriptive inference $\infcirc$ is defined by:
$\Pi \infcirc \Gamma$ if and only if $\Gamma$ is satisfied by all
minimal models of $\Pi$:

\[
\Pi \infcirc \Gamma
\mbox{ ~~ if and only if ~~ }
\circ(\Pi) \subseteq \mod(\Gamma)
\]

\end{definition}

Equivalence in propositional logic can be defined in two
equivalent ways: either by equality of the models or by
equality of the sets of entailed formulae. These two
definitions of equivalence coincide for circumscriptive
inference as well. We define $\equiv_M$ as follows: $\Pi
\equiv_M \Gamma$ if and only if $\circ(\Pi) =
\circ(\Gamma)$. Redundancy of a clause is defined as
follows.

\begin{definition}

A clause $\gamma \in \Pi$ is $\circ$-redundant in the \cnf\
formula $\Pi$ if and only if $\Pi \backslash \{\gamma\}
\equiv_M \Pi$.

\end{definition}

A formula is redundant if some of its clauses can be removed
without changing its semantics.

\begin{definition}

A formula is $\circ$-redundant if it is
$\equiv_M$-equivalent to one of its proper subsets.

\end{definition}

A formula is therefore redundant if some clauses can be
removed from it while preserving equivalence. In the next
section we show that a formula is $\circ$-redundant if and
only if it contains a $\circ$-redundant clause, that is,
circumscription has the local redundancy property.

\subsection{Clause-Redundancy vs. Formula-Redundancy}

Propositional logic has the local redundancy property.
Showing why is interesting for comparison with logics not
allowing the same proof to be used. If $\Pi$ does not
contain a redundant clause, then $\Pi \backslash \{\gamma\}
\not\equiv \Pi$ for any clause $\gamma \in \Pi$. Therefore,
$\mod(\Pi) \not= \mod(\Pi \backslash \{\gamma\})$. Since
$\Pi \backslash \{\gamma\}$ is a subset of $\Pi$, we have
$\mod(\Pi) \subseteq \mod(\Pi \backslash \{\gamma\})$ in
general and $\mod(\Pi) \subset \mod(\Pi \backslash
\{\gamma\})$ in this case. If $\Pi' \subset \Pi$ then $\Pi'
\subseteq \Pi \backslash \{\gamma\}$ for a clause $\gamma$.
Therefore, $\mod(\Pi) \subset \mod(\Pi \backslash
\{\gamma\}) \subseteq \mod(\Pi')$, which proves that $\Pi$
and $\Pi'$ are not equivalent.

This proof does not work for circumscription because the set
of minimal models of a formula can grow or shrink in
response to a clause deletion. In principle, $\Pi$ and $\Pi
\backslash \{\gamma\}$ might have different sets of minimal
models and yet $\Pi$ and $\Pi' \subset \Pi \backslash
\{\gamma\}$ have the same minimal models. We show that this
is not possible. The proof is based on the following simple
result about quasi-orders (reflexive and transitive
relations.)

\begin{lemma}
\label{quasi-min}

If $\leq$ is a quasi-order (a reflexive and transitive
relation) and  $A$ and $B$ are two finite sets such that $A
\subseteq B$ and $\min_\leq(A) \not= \min_\leq(B)$, then
$\min_\leq(B) \backslash A$ is not empty.

\end{lemma}

\proof Since $\min_\leq(A) \not= \min_\leq(B)$, then either
$\min_\leq(A) \backslash \min_\leq(B)$ or $\min_\leq(B)
\backslash \min_\leq(A)$ is not empty. We consider these two
cases separately.

Let $x \in \min_\leq(B) \backslash \min_\leq(A)$. We prove
that $x \not\in A$. Since $x$ is minimal in $B$, there is no
element of $y \in B$ such that $y < x$. Since $A \subseteq
B$, the same holds for every element of $A$ in particular.
As a result, if $x \in A$ then $x \in \min_\leq(A)$,
contradicting the assumption.

Let us instead assume that $\min_\leq(A) \backslash
\min_\leq(B)$ is not empty. Let $x \in \min_\leq(A)
\backslash \min_\leq(B)$. Since $x \in A$, it holds $x \in
B$. Since $x \in B$, $x \not\in \min_\leq(B)$, and $B$ is a
finite set, there exists $y \in \min_\leq(B)$ such that $y <
x$. Since $x$ is minimal in $A$, we have that $y \not\in
A$.~\qed

The order $\subseteq$ on propositional models is a
quasi-order. As a result, if $A$ and $B$ are two sets of
models such that $A \subseteq B$ and the set of minimal
elements of $A$ and $B$ are different, then $B$ has a
minimal element that is not in $A$. When applied to
circumscription, this result tells that a formula can be
non-equivalent to a stronger one only because of a minimal
model that is not a model of the stronger formula. In the
other way around, if a formula is weakened, the set of
minimal models either remains the same or acquires a new
element.

\begin{theorem}

If $\mod(\Pi) \subseteq \mod(\Pi') \subseteq \mod(\Pi'')$
and $\circ(\Pi) \not= \circ(\Pi')$ then $\circ(\Pi) \not=
\circ(\Pi'')$.

\end{theorem}

\proof Let us assume that $\circ(\Pi) = \circ(\Pi'')$. Since
$\circ(\Pi) \not= \circ(\Pi')$, we have $\circ(\Pi') \not=
\circ(\Pi'')$. Since $\subseteq$ on propositional models is
a quasi-order, Lemma~\ref{quasi-min} applies: there exists
$M$ such that $M \in \circ(\Pi'')$ and $M \not\in
\mod(\Pi')$. Since $\mod(\Pi) \subseteq \mod(\Pi')$, the
latter implies $M \not\in \mod(\Pi)$. As a result, $M
\not\in \circ(\Pi)$. Since $M \in \circ(\Pi'')$, we have
$\circ(\Pi) \not= \circ(\Pi'')$.~\qed

The redundancy of a formula and the presence of a redundant
clause in the formula are related by the following theorem,
which is an application of the above to the case in which
$\Pi'=\Pi \backslash \{\gamma\}$ and $\Pi''$ is a subset of
$\Pi'$.

\begin{theorem}
\label{circ-one}

A \cnf\  formula $\Pi$ is \circ-redundant if and only if it
contains a \circ-redundant clause.

\end{theorem}

\proof The ``if'' direction is obvious: if $\gamma$ is
redundant in $\Pi$, then $\Pi \backslash \{\gamma\} \equiv_M
\gamma$, and $\Pi \backslash \{\gamma\}$ is therefore a
strict subset of $\Pi$ that is equivalent to it. The ``only
if'' direction is a consequence of the above theorem. Assume
that $\circ(\Pi \backslash \{\gamma\}) \not= \circ(\Pi)$
holds for every $\gamma \in \Pi$. Let us consider $\Pi''
\subset \Pi$: we prove that $\circ(\Pi'') \not= \circ(\Pi)$.
Since $\Pi'' \subset \Pi$, there exists $\gamma \in \Pi''
\backslash \Pi$. Consider one such clause $\gamma$. Since
$\Pi \subset \Pi \backslash \{\gamma\} \subseteq \Pi''$, we
have that $\mod(\Pi'') \subseteq \mod(\Pi \backslash
\{\gamma\}) \subset \mod(\Pi)$. We are thus in the
conditions to apply the above theorem: since $\circ(\Pi
\backslash \{\gamma\}) \not= \circ(\Pi)$, we have that
$\circ(\Pi'') \not= \circ(\Pi)$. Therefore, $\Pi$ is not
equivalent to any of its proper subsets.~\qed

This theorem shows that circumscription, although
nonmonotonic, has the local redundancy property.

\subsection{Redundant Clauses}

The following lemma characterizes the clauses that are
redundant in a formula.

\begin{lemma}
\label{minimal-negation}

The following three conditions are equivalent:

\begin{enumerate}

\item the clause $\gamma \in \Pi$ is $\circ$-redundant in
$\Pi$;

\item for each $M \in \mod(\Pi \backslash \{\gamma\} \cup
\neg \gamma)$ there exists $M' \in \mod(\Pi)$ such that $M'
\subset M$;

\item for each $M \in \mod(\Pi \backslash \{\gamma\} \cup
\neg \gamma)$ there exists $M' \in \mod(\Pi \backslash
\{\gamma\})$ such that $M' \subset M$.

\end{enumerate}

\end{lemma}

\proof The models of $\Pi \backslash \{\gamma\}$ that are
not models of $\Pi$ are exactly the models of $\Pi
\backslash \{\gamma\} \cup \neg \gamma$. The two formulae
$\Pi$ and $\Pi \backslash \{\gamma\}$ are
$\infcirc$-equivalent if none of these models (if any) is
minimal, that is, all these models contain other models of
$\Pi$. In other words, $\gamma$ is redundant if and only if
every model of $\Pi \backslash \{\gamma\} \cup \neg \gamma$
contains a model of $\Pi$.

The fact that we can check $M' \in \mod(\Pi \backslash
\{\gamma\})$ instead of $M' \in \mod(\Pi)$ follows from the
fact that $\mod(\Pi \backslash \{\gamma\})$ is composed of
all models of $\Pi$ and all models of $\Pi \backslash
\{\gamma\} \cup \neg \gamma$. Consider a model $M$ that is a
minimal model of $\Pi \backslash \{\gamma\} \cup \neg
\gamma$. The condition $M' \subset M$ implies that $M'$ is
not a model of $\Pi \backslash \{\gamma\} \cup \neg \gamma$,
and is therefore a model of $\Pi$. By transitivity, the
condition that there exists $M' \in \mod(\Pi)$ such that $M'
\subset M$ holds for all models of $\Pi \backslash
\{\gamma\} \cup \neg \gamma$.~\qed

Computationally, checking the second or third condition of
this lemma can be done by checking whether for all $M \in
\ldots$ there exists $M' \in \ldots$ such that a simple
condition is met. As a result, the problem is in \P{2}. For
positive clauses, checking $\circ$-redundancy is easier, as
it amounts to checking classical redundancy.

\begin{lemma}
\label{positive-irredundant}

A positive clause is $\circ$-redundant in $\Pi$ if and only
if it is classically redundant in $\Pi$.

\end{lemma}

\proof If a clause is redundant in $\Pi$ it is also
$\circ$-redundant in $\Pi$. Let us now prove the converse:
assume that $\gamma$ is a positive clause that is
$\circ$-redundant in $\Pi$. By Lemma~\ref{minimal-negation},
every model of $\Pi \backslash \{\gamma\} \cup \neg \gamma$
contains a model of $\Pi$, which is the same as $\Pi
\backslash \{\gamma\} \cup \{\gamma\}$. Let $\gamma =
x_{i_1} \vee \cdots \vee x_{i_k}$. Since $\gamma$ is $\circ$
redundant, each model of $\Pi \backslash \{\gamma\} \cup
\{\neg x_{i_i}, \ldots, \neg x_{i_k}\}$ contains at least a
model of $\Pi \backslash \{\gamma\} \cup \{x_{i_i} \vee
\cdots \vee x_{i_k}\}$. All models of the latter formula
contain at least a variable among $x_{i_i}, \ldots, x_{i_k}$
while no models of the former contain any of them.
Therefore, no model of the first formula contains a model of
the second. Therefore, the condition can be true only if
$\Pi \backslash \{\gamma\} \cup \{\neg x_{i_1} \vee \cdots
\vee \neg x_{i_k}\}$ has no models, that is, $\Pi \backslash
\{\gamma\} \models \gamma$: the clause $\gamma$ is
classically redundant in $\Pi$.~\qed

Intuitively, positive clauses only exclude models with all
their literals assigned to false. Therefore, whenever a
positive clause is irredundant \wrt\ $\models$, it is
because such models were not otherwise excluded; therefore,
it is also irredundant \wrt\  minimal models.

According to this argument, it may look like all negative
clauses are redundant because they exclude models with
positive literals, and these models are not minimal. This is
however not the case: a model with some positive literals
might be minimal because no other model of the formula has
less positive literal. Consider, for example, the following
formula:

\[
\Pi=\{\neg x_1 \vee \neg x_2, x_1 \vee x_3, x_2 \vee x_3 \}
\]

The clause $\neg x_1 \vee \neg x_2$, although negative, is
irredundant. Indeed, $\Pi \backslash \{\neg x_1 \vee \neg
x_2\} = \{x_1 \vee x_3, x_2 \vee x_3 \}$, and this formula
has $\{x_1,x_2\}$ and $\{x_3\}$ as its minimal models. The
first one is not a model of $\Pi$ because of the clause
$\neg x_1 \vee \neg x_2$. Therefore, $\neg x_1 \vee \neg
x_2$ is \circ-irredundant in $\Pi$.

Intuitively, a negative clause excludes the possibility of
setting all variables to true, while minimal inference only
{\em tries} to set variables to false. Therefore, removing
the clause may generate a model that have its variables set
to true ($\{x_1,x_2\}$ in the example), but is minimal
because of the values of the other variables ($x_3$ in the
example).

Lemma~\ref{positive-irredundant} can be extended to clauses
containing negative literals via the addition of new clauses
and new variables. To this aim, the following property of
quasi-orders is needed.

\begin{lemma}

If $\leq$ is a quasi-order, $X \in \min_\leq(A)$, $X \in B$,
and $B \subseteq A$, then $X \in \min_\leq(B)$.

\end{lemma}

\proof Since $X \in \min_\leq(A)$, there is no element $Y
\in A$ such that $Y < X$. Since $B \subseteq A$, there is no
element of $B$ with the same property. Since $X$ is an
element of $B$ such that $Y < X$ does not hold for any $Y
\in B$, it holds $X \in \min_\leq(B)$ by definition.~\qed

Applied to formulae: if $M$ is a minimal model of $\Pi$ and
satisfies $\Pi'$, then $M$ is a minimal model of $\Pi
\cup \Pi'$.

\begin{lemma}
\label{negative-irredundant}

A clause $\gamma$ is classically redundant in $\Pi$ if and
only if it is $\circ$-redundant in $\Pi \cup \{x \vee x' ~|~
\neg x \in \gamma\}$.

\end{lemma}

\proof If $\gamma$ is redundant in $\Pi$ then $\Pi
\backslash \{\gamma\} \models \gamma$ and therefore $\Pi
\cup \{x \vee x' ~|~ \neg x \in \gamma\} \backslash
\{\gamma\} \models \gamma$. Since $\gamma$ is redundant in
$\Pi \cup \{x \vee x' ~|~ \neg x \in \gamma\}$, it is also
$\circ$-redundant.

Let us now assume that $\gamma$ is irredundant in $\Pi$,
that is, $\Pi \backslash \{\gamma\} \cup \neg \gamma$ has
some models. Let $M$ be one such model. Since this model
satisfies $\neg \gamma$, it assigns false to any variable
$x$ such that $x \in \gamma$ and true to any variable $x$
such that $\neg x \in \gamma$. Extending $M$ to assign false
to all variables $x'$, this model also satisfies $\Pi \cup
\{x \vee x' ~|~ \neg x \in \gamma\} \backslash \{\gamma\}
\cup \neg \gamma$.

We show that $M$ cannot contain a model of $\Pi \cup \{x
\vee x' ~|~ \neg x \in \gamma\}$. This model assigns false
to all $x \in \gamma$ and also false to all $x'$ such that
$\neg x \in \gamma$. On the other hand, $\{\gamma\} \cup \{x
\vee x' ~|~ \neg x \in \gamma\}$ entails the clause $\bigvee
\{x ~|~ x \in \gamma\} \vee \bigvee \{x' ~|~ \neg x \in
\gamma\}$; this can be proved for example by iteratively
resolving upon all literals $x$ such that $\neg x \in
\gamma$. As a result, no model of $\Pi \cup \{x \vee x' ~|~
\neg x \in \gamma\}$ has a model that assign false to all $x
\in \gamma$ and all $x'$ such that $\neg x \in \gamma$.
Since this is instead done by $M$, it follows that no model
of $\Pi \cup \{x \vee x' ~|~ \neg x \in \gamma\}$ is
contained in $M$.~\qed

Note that the clauses $x \vee x'$ are not necessarily
$\circ$-irredundant in the considered formula. On the other
hand, Lemma~\ref{positive-irredundant} can be applied to
them: they are $\circ$-redundant if and only if they are
classically redundant.

\subsection{Complexity of Clause Redundancy}

Let us now turn to the hardness of the problem of checking
the redundancy of a clause in a formula. We first show a
reduction that proves the hardness of the problem of
redundancy of a clause and then show how this result can be
used to prove that the problem of redundancy of a formula
has the same complexity.

\begin{theorem}
\label{circ-reduction}

Checking the \circ-redundancy of a clause in a formula is
$\P{2}$-complete.

\end{theorem}

\proof Lemma~\ref{minimal-negation} proves that the
redundancy of a clause in a formula can be checked by
solving a $\forall\exists$QBF (for all $M$... there exists
$M'$...), and is therefore in \P{2}.

Let us now show hardness. We show that the QBF formula
$\forall X \exists Y . \Gamma$, where $\Gamma = \{ \delta_1,
\ldots, \delta_m \}$ and $n=|Y|$, is valid if and only if
$\gamma$ is $\circ$-redundant in $\Pi$, where:

\begin{eqnarray*}
\Pi &=&
\{ x_i \vee p_i \}
\cup
\{ \neg a \vee y_i \}
\cup
\{ a \vee \delta_i ~|~ \delta_i \in \Gamma \}
\cup
\{\gamma\}
\\
\gamma &=&
\neg a \vee
\neg y_1 \vee \cdots \vee \neg y_n
\end{eqnarray*}

The clause $\gamma$ is \circ-redundant in $\Pi$ if and only
if all minimal models of $\Pi \backslash \{\gamma\} \cup
\neg \gamma$ contain some models of $\Pi$. The following
equivalences holds:

\begin{eqnarray*}
\Pi 
&\equiv&
\{x_i \vee p_i\} \cup \{\neg a\} \cup \Gamma
\\
\Pi \backslash \{\gamma\} \cup \neg \gamma
&\equiv&
\{ x_i \vee p_i \}
\cup
\{a, y_1, \ldots, y_n \}
\end{eqnarray*}

The first equivalence holds because $\{\neg a \vee y_i\}
\cup \{\neg a \vee \neg y_1 \vee \cdots \vee \neg y_n \}$ is
equivalent to $\neg a$, as can be checked by resolving upon
each $y_i$ in turn. The second equivalence holds because
$\neg \gamma = \{a, y_1, \ldots, y_n \}$ and this set
implies all clauses $\neg a \vee y_i$ and $a \vee \delta_i$.

The formula $\Pi \backslash \{\gamma\} \cup \neg \gamma
\equiv \{ x_i \vee p_i \} \cup \{a, y_1, \ldots, y_n \}$ has
a minimal model for each truth evaluation $\omega_X$ over
the variables $x_i$:

\[
I_{\omega_X} = 
\omega_X \cup \{p_i ~|~ x_i \not\in \omega_X \}
\cup
\{a\}
\cup
\{y_i ~|~ 1 \leq i \leq n\}
\]

We show that the model $I_{\omega_X}$ contains a model of
$\Pi$ if and only if $\Gamma|_{\omega_X}$ is satisfiable. By
Lemma~\ref{minimal-negation}, the redundancy of $\gamma$
corresponds to this condition being true for all possible
models of $\Pi' \cup \neg \gamma$. This would therefore
prove that the QBF is valid if and only if $\gamma$ is
redundant in $\Pi$.

Since $\Pi \equiv \{x_i \vee p_i\} \cup \{\neg a\} \cup
\Gamma$, if $\Gamma$ has a model with a given value of
$\omega_X$ then $\Pi$ has a model that is strictly contained
in $I_{\omega_X}$: add to the satisfying assignment of
$\Gamma$ the setting of every $p_i$ to the opposite of $x_i$
and $a$ to false.

On the converse, if $\Pi$ contains a model that is strictly
contained in $I_{\omega_X}$, this model must have exactly
the same value of $X \cup P$ because $\Pi$ contains $x_i
\vee p_i$ and either $x_i$ or $y_i$ is false in
$I_{\omega_X}$. On the other hand, this model of $\Pi$ must
also set $a$ to false and satisfy $\Gamma$, thus showing
that there exists an assignment extending $\omega_X$ and
satisfying $\Gamma$.\qed

\subsection{Complexity of Formula Redundancy}

In order to characterize the complexity of the problem of
checking the \circ-redundancy of a formula, we use the fact
that a formula is \circ-redundant if and only if it contains
a \circ-redundant clause by Theorem~\ref{circ-one}. In
particular, Lemma~\ref{circ-reduction} shows that the
problem of checking the $\circ$-redundancy of a clause
$\gamma$ in $\Pi$ is \P{2}-hard. In order for this result to
be used as a proof of hardness for the problem of
\circ-redundancy of formulae, we need to modify the formula
$\Pi$ in such a way all its clauses but $\gamma$ are made
\circ-irredundant. This is the corresponding of Lemma~4 of
the paper of redundancy of propositional \cnf\  formulae
\cite{libe-05}, which has been useful because it allows to
``localize'' problems about redundancy.

\begin{lemma}
\label{making-irredundant}

For every consistent formula $\Pi$ and $\Pi' \subseteq \Pi$,
the only $\circ$-redundant clauses of $I(\Pi,\Pi')$ are the
clauses $\neg s \vee \neg t \vee \gamma_i$ such that
$\gamma_i \in \Pi'$ and $\gamma_i$ is $\circ$-redundant in
$\Pi$.

\begin{eqnarray*}
I(\Pi,\Pi') &=&
\{s \vee t\}
\cup
\{s \vee a, t \vee b\}
\cup
\\&&
\{\neg s \vee t \vee c_i \vee d_i\}
\cup
\{\neg s \vee \neg c_i\}
\cup
\\&&
\{\neg t \vee c_i \vee \gamma_i ~|~ \gamma_i \in \Pi \backslash \Pi'\}
\cup
\{s \vee \neg t \vee x \vee x' ~|~ x \in \var(\Pi) \} \cup
\\&&
\{\neg s \vee \neg t \vee \gamma_i ~|~ \gamma_i \in \Pi'\}
\end{eqnarray*}

\end{lemma}

\proof There are four possible assignment to the variables
$s$ and $t$. Since the models of $I(\Pi,\Pi')$ can be
partitioned into the models of $I(\Pi,\Pi') \cup \{\neg s,\neg
t\}$, $I(\Pi,\Pi') \cup \{s,\neg t\}$, $I(\Pi,\Pi') \cup
\{\neg s,t\}$, and $I(\Pi,\Pi') \cup \{s,t\}$, the minimal
models of $I(\Pi,\Pi')$ are necessarily some of the minimal
models of these formulae.

In the table below we show what remains of $I(\Pi,\Pi')
\backslash \{s \vee t\}$ in each of the four possible
assignment to $s$ and $t$ after removing entailed clauses
and false literals. We also show  the minimal models of the
resulting formulae.

\begin{center}
\begin{tabular}{llll}
assignment & subformula & minimal models\\
\hline
$\{\neg s, \neg t\}$
&
$\{a,b\}$
&
$\{a,b\}$
\\
$\{s, \neg t\}$
&
$\{b\} \cup \{c_i \vee d_i\} \cup \{\neg c_i\}$
&
$\{s,b\} \cup \{d_i\}$
\\
$\{\neg s, t\}$
&
$\{a\} \cup \{c_i \vee \gamma_i ~|~ \gamma_i \in \Pi \backslash \Pi'\} \cup$
&
$\{t,a\} + \mbox{some subsets of } (C \cup X \cup X')$
\\
& $\{x \vee x' ~|~ x \in \var(\Pi)\}$
\\
$\{s,t\}$
&
$\{\neg c_i\} \cup
\{c_i \vee \gamma_i ~|~ \gamma_i \in \Pi \backslash \Pi'\} \cup \Pi'$
&
$\{s,t\} + \mbox{a minimal model of } \Pi$
\end{tabular}
\end{center}

The four subformulae are all satisfiable. Moreover, no
minimal model of one is contained in the minimal models of
the other ones because of either the values of $\{s,t\}$ and
$\{a,b\}$. As a result, the minimal models of
$I(\Pi,\Pi')\backslash \{s \vee t\}$ are exactly the minimal
models of the four subformulae. The clause $s \vee t$ is
irredundant because its addition deletes the minimal model
$\{a,b\}$. The minimal models of $I(\Pi,\Pi')$ are therefore
exactly the minimal models of the remaining three
subformulae.

We show that the remaining clauses but the ones derived from
$\Pi'$ are irredundant. This is shown by removing a clause
from the set and showing that some of the minimal models of
a subformula can be removed some elements. Since the minimal
models of these three subformulae are exactly the minimal
models of $\Pi$, this is a proof that the clause is
irredundant.

\begin{enumerate}

\item
The clauses $s \vee a$ and $t \vee b$ are irredundant
because their removal would allow $a$ and $b$ to be set to
false in the minimal models of the third and second
subformula, respectively.

\item
The clauses $\neg s \vee t \vee c_i \vee d_i$ and $\neg s
\vee \neg c_i$ are irredundant because their removal would
allow $d_i$ to be set to false in the minimal model of the
second subformula.

\item
The clauses $\neg t \vee c_i \vee \gamma_i$ and $s \vee \neg
t \vee x \vee x'$ require a longer analysis. In the third
assignment, $I(\Pi,\Pi')$ becomes:

\[
C=
\{a\} \cup
\{c_i \vee \gamma_i\} \cup
\{x \vee x' ~|~ x \in \var(\Pi)\}
\]

The clauses $x \vee x'$ are positive. By
Lemma~\ref{positive-irredundant}, they are $\circ$-redundant
if and only if they are redundant. In turn, they are not
redundant because $\{a\} \cup \{c_i\} \cup \{y ~|~ y \not=
x\}$ is a model of all clauses but $x \vee x'$.

Since $c_i$ occurs positive in $c_i \vee \gamma_i$,
Lemma~\ref{negative-irredundant} ensures that this clause is
$\circ$-redundant in $C$ if and only if it is redundant in
$C \backslash \{x \vee x' ~|~ \neg x \in \gamma_i\}$. This
is false because the removal of $c_i \vee \gamma_i$ creates
the following new model:

\[
M = \{c_j ~|~ j \not= i\} \cup \{x'\}
\cup \{x ~|~ \neg x \in \gamma_i\}
\]

This model $M$ satisfies $C \backslash \{x \vee x' ~|~ \neg
x \in \gamma_i\} \backslash \{c_i \vee \gamma_i\}$: all
clauses $c_j \vee \gamma_j$ are satisfied because $c_j \in
M$ and all clauses $x \vee x'$ are satisfied because $x' \in
M$. On the other hand, $M$ does not satisfy $c_i \vee
\gamma_i$ because it assigns all its literals to false.

\end{enumerate}

The only clauses that can therefore be redundant are those
corresponding to the clauses of $\Pi'$. In particular, these
clauses only occur in the fourth subformula, which is
equivalent to $\{c\} \cup \{\neg c_i\} \cup \Pi$. A clause
$\neg s \vee \neg t \vee \gamma_i$ with $\gamma_i \in \Pi'$
is therefore $\circ$-redundant in $I(\Pi,\Pi')$ if and only
if $\gamma_i$ is $\circ$-redundant in $\Pi$.~\qed

More precisely, this theorem shows a way to make the clauses
of $\Pi'$ {\em necessary}, that is, contained in all
equivalent subsets of $\Pi$. The theorem allows to
characterize the complexity of formula \circ-redundancy.

\begin{theorem}

The problem of $\circ$-redundancy is \P{2}-complete.

\end{theorem}

\proof By Theorem~\ref{circ-one}, $\Pi$ is redundant if and
only if it contains a redundant clause. Therefore, we have
to solve a linear number of problems in \P{2}. Since these
problems can be solved in parallel, the whole problem is in
\P{2}.

Hardness is proved by reduction from the problem of
$\circ$-redundancy of a single clause. By
Lemma~\ref{making-irredundant}, a clause $\gamma$ is
\circ-redundant in $\Pi$ if and only if $\neg s \vee \neg t
\vee \gamma$ is \circ-redundant in $I(\Pi,\{\gamma\})$ and
all other clauses of $I(\Pi,\{\gamma\})$ are
irredundant.~\qed

%%%%%%%%%%%%% end:circ %%%%%%%%%%%%%%
 %

\let\sectionnewpage=\newleaf

%%%%%%%%%%%%% begin:default %%%%%%%%%%%%%%
\section{Default Logic}

\def\x{ispell ignore this part \[}

\def\ext{{\rm Ext}}
\def\extd{\ext_D}

\def\infdef{\models_D}

\def\prec{{\rm prec}}
\def\just{{\rm just}}
\def\cons{{\rm cons}}

\def\x{\]}

A default theory is a pair $\l D,W \r$, where $W$ is formula
and $D$ is a set of default rules, each rule being in the
form:

\[
\frac{\alpha:\beta}
{\gamma}
\]

The formulae $\alpha$, $\beta$, and $\gamma$ are called the
precondition, the justification, and the consequence of the
default, respectively. In this paper, we assume that $W$ is
a \cnf\  finite formula (a finite set of clauses) and that
the set of variables and defaults are finite. We also assume
that each default has a single justification, rather than a
set of justifications. Given a default $d =
\frac{\alpha:\beta}{\gamma}$, its parts are denoted by
$\prec(d)=\alpha$, $\just(d)=\beta$, and $\cons(d)=\gamma$.

We use the operational semantics of default logics
\cite{anto-sper-94,anto-99,froi-meng-92,froi-meng-94}, which
is based on sequences of defaults with no duplicates. If
$\Pi$ is such a sequence, we denote by $\Pi[d]$ the sequence
of defaults preceeding $d$ in $\Pi$, and by $\Pi \cdot [d]$
the sequence obtained by adding $d$ at the end of $\Pi$. We
extend the notation from defaults to sequences, so that
$\prec(\Pi)$ is the conjunction of all preconditions of the
defaults in $\Pi$, $\just(\Pi)$ is the conjunction of all
justifications, and $\cons(\Pi)$ is the conjunction of all
consequences.

Implication is denoted by $\models$, $\top$ indicates
(combined) consistency, and $\bot$ indicates inconsistency.
For example, $A \top B$ means that $A \wedge B$ is consistent,
while $A \bot B$ means that $A \wedge B$ is inconsistent.

Default logic can be defined in terms of the {\em selected
processes}, that are the sequences of defaults that are
considered applicable by the semantics~\cite{anto-99}. A
sequence of defaults $\Pi$ is a process if $W \cup
\cons(\Pi[d]) \models \prec(d)$ holds for any $d \in \Pi$. A
default $d$ is {\em locally applicable} in a sequence $\Pi$
if $\cons(\Pi) \cup W \models \prec(d)$ and $\cons(\Pi) \cup
W \top \just(d)$. Global applicability also requires
$\cons(\Pi) \cup W \top \just(\Pi \cdot [d])$. Each
semantics defines the sequences of defaults that are applied
in a particular theory. Formally, the definitions are as
follows:

\begin{description}

\item[Reiter:] a process $\Pi$ is selected if $\cons(\Pi)
\cup W \top \just(d)$ for each $d \in \Pi$ and no
default $d' \not\in \Pi$ is locally applicable in $\Pi$;

\item[Justified:] a process is selected if it is a maximal
process such that $\cons(\Pi) \cup W \top \just(d)$ for each
$d \in \Pi$;

\item[Constrained:] a process is selected if it is a maximal
process such that $\cons(\Pi) \cup W \top \just(\Pi)$;

\item[Rational:] a process is selected if $\cons(\Pi) \cup W
\models \prec(d)$ and no default $d' \not\in \Pi$ is
globally applicable in $\Pi$.

\end{description}

The conditions on selected processes can be all broken in
two parts: success (the consistency condition) and closure
(the non-extendibility of the process). For example, for
constrained default logic the condition of success is
$\cons(\Pi) \cup W \top \just(d)$ and the condition of
closure is that $\Pi \cdot [d]$ is not successful for any
$d' \not\in \Pi$.

Remarkably, the conditions above only mention the background
theory $W$ in conjunction with $\cons(\Pi)$, that is, $W$
only occurs in subformulae of the form $W \cup \cons(\Pi)$.
The only conditions for which this is not true is that of
$\Pi$ being a process.

If $\Pi$ is a selected process of $\l D,W \r$, the formula
$Cn(W \cup \cons(\Pi))$ is an extension of $\l D,W \r$. We
denote by $\ext(\l D,W \r)$ or $\extd(W)$ the set of all
formulae that are equivalent to an extension of $\l D,W \r$.
Including formulae that are equivalent to the extensions in
this set allows to write $E \in \extd(W)$ to denote the
equivalence of $E$ with an extension of $\l D,W \r$.

A default theory $\l D,W \r$ entails a formula $W'$ if and
only if $E \models W'$ for every $E \in \extd(W)$. This
condition is equivalent to $\vee \extd(W) \models W'$; as a
result, the set of all consequences of a default theory is
equivalent to $\vee \extd(W)$. The condition that $\l D,W
\r$ entails $W'$ is denoted by $\l D,W \r \models W'$ or $W
\models_D W'$. The latter notation emphasizes that every
fixed set of defaults $D$ induces a nonmonotonic inference
operator $\models_D$.

Some semantics of default logic do not assign any extension
to some theories. In this paper, we try to derive the
hardness results using only theories having extensions.

\subsection{Equivalence in Default Logics}

The monotonic inference operator $\models_D$ induced by a
set of default $D$ is a consequence relation. Therefore, the
definitions of redundancy of a clause and of a formula for
$\models$ and $\infcirc$ can be given for $\models_D$ as
well: a clause $\gamma$ of a formula $W$ is redundant \wrt\
default $D$ if and only if $W$ and $W \backslash \{\gamma\}$
are equivalent; a formula $W$ is redundant if there exists
$W' \subset W$ that is equivalent to it.

Both definitions are based on equivalence of two formulae,
and in particular the equivalence of a formula with one of
its proper subsets. In this section, we show that three
different form of equivalence can be defined; we compare
them in general and in the particular case of equivalence of
a formula with one of its proper subsets. The first form of
equivalence is based on entailment.

\begin{definition}
[Entailment and Mutual Equivalence]

For a given set of defaults $D$, formula $W$ entails $W'$,
denoted by $W \infdef W'$, if $\vee \extd(W) \models W'$.
These two formulae are mutual equivalent, denoted by $W
\equiv^m_D W'$, if $W \infdef W'$ and $W' \infdef W$.

\end{definition}

In classical logic, this definition of equivalence is the
same as $W$ and $W'$ having the same set of consequences and
the same set of models. In default logic, this is not the
case. We define the equivalence based on the set of
consequences as follows.

\begin{definition}
[Consequence-Entailment and Consequence-Equivalence]

For a given set of defaults $D$, formula $W$
consequence-entails $W'$, denoted $W \infdef^c W'$, if $\vee
\extd(W) \models \vee \extd(W')$. These two two formulae are
consequence-equivalent, denoted $W \equiv^c_D W'$, if $\vee
\extd(W) \equiv \vee \extd(W')$.

\end{definition}

Note that the comparison of the two formulae is based on a
fixed set of defaults $D$. A more stringent condition of
equivalence of two defaults theories is that of having the
same extensions.

\begin{definition}
[Faithful Entailment and Faithful Equivalence]

For a given set of defaults $D$, a formula $W$ faithfully
entails $W'$, denoted $W \infdef^e W'$, if $\extd(W)
\subseteq \extd(W')$. These two formulae are faithfully
equivalent, denoted $W \equiv^e_D W'$, if $\extd(W) =
\extd(W')$.

\end{definition}

For all three definition, equivalence is the same as each
formula implying the other one. We are especially interested
into $\equiv_D^c$ and $\equiv_D^e$, that is, equality of
consequences and equality of extensions. Mutual equivalence
has been defined for technical reasons. Redundancy in
default logic is defined as follows.

\begin{definition}
[Redundancy of a Clause]

For a given set of defaults $D$, a clause $\gamma$ is
redundant in a formula $W$ according to equivalence
$\equiv_D^x$ if $W \equiv_D^x W \backslash \{\gamma\}$.

\end{definition}

\begin{definition}
[Redundancy of a Formula]

For a given set of defaults $D$, a formula $W$ is redundant
according to equivalence $\equiv_D^x$ if there exists $W'
\subset W$ such that $W \equiv_D^x W'$.

\end{definition}

In both cases, we are comparing for equivalence a formula
and one of its subsets. In the following section we study
the equivalence of $W'$ and $W$ when $W' \subseteq W$.

\subsubsection{Correspondence, General}

We now compare the three forms of equivalence defined above.
The following chain of implications is easy to prove:

\[
W' \infdef^e W
~~ \Rightarrow ~~
W' \infdef^c W
~~ \Rightarrow ~~
W' \infdef W
\]

The latter implication is proved by the following lemma.

\begin{lemma}
\label{derive-consequences}

If $W' \infdef^c W$ then $W' \infdef W$.

\end{lemma}

\proof By assumption, $\vee \extd(W') \models \vee
\extd(W)$. Since every extension of $\l D,W \r$ entails $W$,
we have $\vee \extd(W) \models W$. As a result, $\vee
\extd(W') \models W$, which is by definition $W' \infdef
W$.~\qed

Redundancy is defined in terms of equivalence of two
formulae, one contained in the other. As a result, it makes
sense to study the conditions of equivalence in the
particular case in which $W' \subseteq W$. We prove that the
above chain of implication can be wrapped around in this
case, thus proving that the three conditions are equivalent.

\begin{lemma}
\label{derive-theory}

If $W' \subseteq W$ and $W' \infdef W$, then $W' \infdef^e
W$.

\end{lemma}

\proof Let $\Pi$ be a selected process of $\l D,W' \r$. We
prove that it is also a selected process of $\l D,W \r$.
Since $W' \infdef W$, the formula $W$ is entailed by every
extension in $\extd(W')$. In particular, $W' \cup \cons(\Pi)
\models W$. Therefore, $W' \cup \cons(\Pi) \equiv W \cup
\cons(\Pi)$. As a result, all conditions (such as success
and closure) where $W'$ only occurs in the subformula $W' \cup
\cons(\Pi)$ are not changed by replacing $W'$ with $W$. This
is in particular true for all considered conditions of
success and closure.

The only condition that mentions the background theory not
in conjunction with $\cons(\Pi)$ is the condition of a
sequence being a process: $\Pi$ is a process of $\l D,W' \r$
if and only if $W' \cup \cons(\Pi[d]) \models \prec(d)$ for
any $d \in \Pi$. The same condition is however true for $W$
because $W' \subseteq W$ implies $W \models W'$.~\qed

The following is a consequence of the above.

\begin{corollary}
\label{entail-same}

If $W' \subseteq W$, then:

\[
W' \infdef W
~~ \Leftrightarrow ~~
W' \infdef^c W
~~ \Leftrightarrow ~~
W' \infdef^e W
\]

\end{corollary}

\subsubsection{Non-Correspondence, General}

The last corollary proves that the three definitions of
entailment from $W'$ to $W$ are equivalent if $W' \subseteq
W$. The same does not hold for equivalence, and therefore
does not hold for entailment from $W$ to $W'$.

\begin{theorem}
\label{mutual-not-consequence}

There exists $D$, $W$, and $W' \subset W$ such that
$W' \equiv^m_D W$ and $W' \not\equiv^c_D W$

\end{theorem}

\proof Since $W' \infdef W$, every extension of $\l D,W' \r$
implies $W$. A wrong proof of $W' \equiv^e_D W$ could then
be based on the fact that, once $W$ is derived from $\l D,W'
\r$ applying some defaults, we can proceed by applying the
defaults of an arbitrary process of $\l D,W \r$.

\

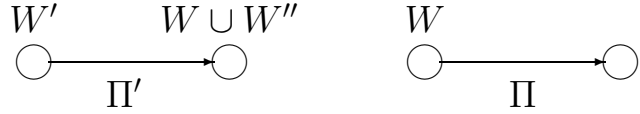
\hfil
%%%%%%%%%%%%% begin:process-continue.latex %%%%%%%%%%%%%%
\setlength{\unitlength}{4144sp}%
\begingroup\makeatletter\ifx\SetFigFont\undefined%
\gdef\SetFigFont#1#2#3#4#5{%
  \reset@font\fontsize{#1}{#2pt}%
  \fontfamily{#3}\fontseries{#4}\fontshape{#5}%
  \selectfont}%
\fi\endgroup%
\begin{picture}(3850,672)(1300,-565)
{\color[rgb]{0,0,0}\thinlines
\put(1531,-241){\circle{202}}
}%
{\color[rgb]{0,0,0}\put(2701,-241){\circle{202}}
}%
{\color[rgb]{0,0,0}\put(5041,-241){\circle{202}}
}%
{\color[rgb]{0,0,0}\put(3871,-241){\circle{202}}
}%
{\color[rgb]{0,0,0}\put(1621,-241){\vector( 1, 0){990}}
}%
{\color[rgb]{0,0,0}\put(3961,-241){\vector( 1, 0){990}}
}%
\put(1531,-61){\makebox(0,0)[b]{\smash{{\SetFigFont{14}{16.8}{\familydefault}{\mddefault}{\updefault}{\color[rgb]{0,0,0}$W'$}%
}}}}
\put(2071,-511){\makebox(0,0)[b]{\smash{{\SetFigFont{14}{16.8}{\familydefault}{\mddefault}{\updefault}{\color[rgb]{0,0,0}$\Pi'$}%
}}}}
\put(3871,-61){\makebox(0,0)[b]{\smash{{\SetFigFont{14}{16.8}{\familydefault}{\mddefault}{\updefault}{\color[rgb]{0,0,0}$W$}%
}}}}
\put(2701,-61){\makebox(0,0)[b]{\smash{{\SetFigFont{14}{16.8}{\familydefault}{\mddefault}{\updefault}{\color[rgb]{0,0,0}$W \cup W''$}%
}}}}
\put(4456,-511){\makebox(0,0)[b]{\smash{{\SetFigFont{14}{16.8}{\familydefault}{\mddefault}{\updefault}{\color[rgb]{0,0,0}$\Pi$}%
}}}}
\end{picture}%
%%%%%%%%%%%%% end:process-continue.latex %%%%%%%%%%%%%%
 %

\

This figure shows why a process $\Pi'$ of $\l D,W' \r$ and a
process $\Pi$ of $\l D,W \r$ cannot always be concatenated:
while $\Pi'$ allows the derivation of $W$, this process
might also derive another formula $W''$ that makes the
process $\Pi$ inapplicable. An example in which this
situation arises is the following one:

\begin{eqnarray*}
D &=& \{d_1, d_2\} \\
& \mbox{\hbox to 0pt{where:\hss}} \\
&& d_1 = \frac{:a \wedge \neg b}{a \wedge \neg b}\\
&& d_2 = \frac{a:b}{b}
\\
W &=& \{a\}
\\
W' &=& \emptyset
\end{eqnarray*}

The only process of $\l D,W' \r$ is $[d_1]$, which generates
the extension $Cn(a \wedge \neg b)$. This extension entails
$W$, but it also entails $\neg b$. The theory $\l D,W \r$
has also the process $[d_2]$, generating the extension $Cn(a
\wedge b)$. These two processes cannot however be
concatenated, as the consequence $\neg b$ of $d_1$ is
inconsistent with the justification of $d_2$.

Since $W' \subset W$, we have that $W \infdef W'$. In this
example, we also have $W' \infdef W$ because the single
extension of $W'$ entails $W=\{a\}$. However, $W'$ and $W$
have different set of extensions; in particular, $\vee
\extd(W') \equiv a \wedge \neg b$ and $\vee \extd(W) \equiv
a$.~\qed

A similar result can be proved about $\equiv^c_D$ and
$\equiv_D^e$.

\begin{theorem}
\label{consequence-not-faithful}

There exists $D$, $W$, and $W' \subset W$ such that $W'
\equiv_D^c W$ but $W' \not\equiv_D^e W$ in Reiter and
justified default logic.

\end{theorem}

\proof Rather than the counterexample itself, it is
interesting to show how it has been derived. The idea is the
same as that of Theorem~\ref{mutual-not-consequence}: a
process of $\l D,W' \r$ that cannot be concatenated with a
process of $\l D,W \r$. The proof for this case, however, is
complicated by the fact that we assume $W' \equiv_D^c W$,
that is, $\vee \extd(W')$ and $\vee \extd(W)$ are
equivalent. The theories used in
Theorem~\ref{mutual-not-consequence} do not work, as any
other pair of theories having one extension each: in this
cases, indeed, $\vee \extd(W)$ is equivalent to $\extd(W)$.
In order for the counterexample to work, $\l D,W' \r$ must
have multiple processes, each entailing $W$ and some other
formula.

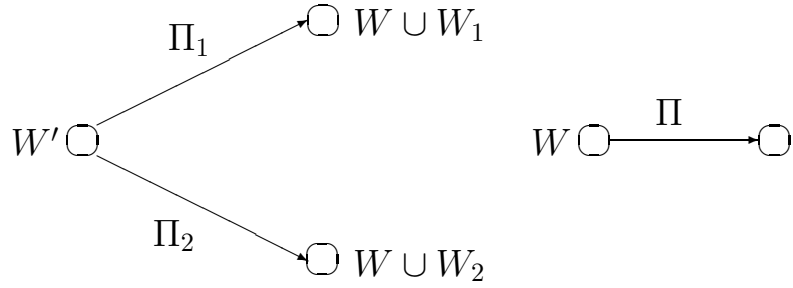
\begin{center}
%%%%%%%%%%%%% begin:implieslarger2.latex %%%%%%%%%%%%%%
\setlength{\unitlength}{4144sp}%
\begingroup\makeatletter\ifx\SetFigFont\undefined%
\gdef\SetFigFont#1#2#3#4#5{%
  \reset@font\fontsize{#1}{#2pt}%
  \fontfamily{#3}\fontseries{#4}\fontshape{#5}%
  \selectfont}%
\fi\endgroup%
\begin{picture}(4835,1690)(394,-1744)
{\color[rgb]{0,0,0}\thinlines
\put(991,-871){\oval(180,180)}
}%
{\color[rgb]{0,0,0}\put(4051,-871){\oval(180,180)}
}%
{\color[rgb]{0,0,0}\put(5131,-871){\oval(180,180)}
}%
{\color[rgb]{0,0,0}\put(2431,-151){\oval(180,180)}
}%
{\color[rgb]{0,0,0}\put(2426,-1581){\oval(180,180)}
}%
{\color[rgb]{0,0,0}\put(1081,-781){\vector( 2, 1){1260}}
}%
{\color[rgb]{0,0,0}\put(1081,-961){\vector( 2,-1){1260}}
}%
{\color[rgb]{0,0,0}\put(4141,-871){\vector( 1, 0){900}}
}%
\put(4501,-781){\makebox(0,0)[b]{\smash{{\SetFigFont{14}{16.8}{\familydefault}{\mddefault}{\updefault}{\color[rgb]{0,0,0}$\Pi$}%
}}}}
\put(856,-961){\makebox(0,0)[rb]{\smash{{\SetFigFont{14}{16.8}{\familydefault}{\mddefault}{\updefault}{\color[rgb]{0,0,0}$W'$}%
}}}}
\put(3916,-961){\makebox(0,0)[rb]{\smash{{\SetFigFont{14}{16.8}{\familydefault}{\mddefault}{\updefault}{\color[rgb]{0,0,0}$W$}%
}}}}
\put(2611,-1681){\makebox(0,0)[lb]{\smash{{\SetFigFont{14}{16.8}{\familydefault}{\mddefault}{\updefault}{\color[rgb]{0,0,0}$W \cup W_2$}%
}}}}
\put(2611,-241){\makebox(0,0)[lb]{\smash{{\SetFigFont{14}{16.8}{\familydefault}{\mddefault}{\updefault}{\color[rgb]{0,0,0}$W \cup W_1$}%
}}}}
\put(1756,-331){\makebox(0,0)[rb]{\smash{{\SetFigFont{14}{16.8}{\familydefault}{\mddefault}{\updefault}{\color[rgb]{0,0,0}$\Pi_1$}%
}}}}
\put(1666,-1501){\makebox(0,0)[rb]{\smash{{\SetFigFont{14}{16.8}{\familydefault}{\mddefault}{\updefault}{\color[rgb]{0,0,0}$\Pi_2$}%
}}}}
\end{picture}%
%%%%%%%%%%%%% end:implieslarger2.latex %%%%%%%%%%%%%%
 %

\end{center}

In order for the counterexample to work, some defaults of
$\Pi$ cannot be applied after $\Pi_1$ or $\Pi_2$ because
their justifications are inconsistent with $W_1$ or $W_2$.
In order for $W \equiv^c_D W'$ to hold, however, $W \cup
\cons(\Pi)$ must be equivalent to $W \cup (W_1 \vee W_2)$.
As a result, every model of $W \cup \cons(\Pi)$ is a model
of $W \cup W_1$ or $W \cup W_2$.

The precondition of the first default of $\Pi$ is entailed
by $W \cup W_1$ and $W \cup W_2$. Since the justifications
of the defaults in $\Pi$ are consistent with the
consequences of $\Pi$, there is a model that is both a model
of $\cons(\Pi) \cup W$ and a model of $\just(d)$ for any $d
\in \Pi$. But this is also a model of $W \cup W_1$ or $W
\cup W_2$. As a result, the default $d$ is applicable in
$\Pi_i$.

This arguments cannot be extended further, however. Indeed,
$\Pi$ may be composed of two defaults, one applicable to $W
\cup W_1$ and one applicable to $W \cup W_2$. This is
possible in a selected process because Reiter and justified
semantics does not enforce joint consistency of
justifications.

A minimal counterexample requires two defaults that can be
applied in $W'$ leading to two disjoint extensions $W \cup
W_1$ and $W \cup W_2$, and two other defaults that can be
applied in sequence from $W$, but not from $W \cup W_1$ or
$W \cup W_2$.

The background theory $W$ of this counterexample is composed
of the four possible models $A$, $B$, $C$, and $D$. We
define $D$ so that $W$ has a process that generates an
extension $E$ having $A$ and $B$ as its models. In order for
the consequences to be the same of those of $W'$, both $A$
and $B$ have to be part of some extensions of $W'$. We
define the defaults so that $W'$ has two processes
generating $A$ and $B$, respectively. Namely, the first
process generates $A$, but its justifications are
satisfiable because the extension contains $C$; the other
one contains $B$, but consistency with justifications are
ensured by the model $D$. This trick is necessary to avoid
these processes to be extended with the defaults that
generate $W'$.

In order to make the discussion more intuitive, we
identify models with terms, and define formulae and
defaults based on terms. We then convert terms into
real formulae. The defaults that are applicable from
$W'$ are defined as follows.

\begin{eqnarray*}
d_1 &=& \frac{:C}{A \vee C} \\
\\
d_2 &=& \frac{:D}{B \vee D}
\end{eqnarray*}

Both $d_1$ and $d_2$ can be applied from $W'$, leading to a
consequence that is inconsistent with the justification of
the other default. Moreover, each extension contains a model
of $\{A,B\}$, as required. For example, $d_1$ produces $A$.
However, the consistency of the extension with the
justification is ensured by the other model $C$. This is
necessary to avoid these defaults to be applicable in the
new extension $E$ of $W$. Let us now define the two defaults
that are applicable from $W$ only and generate this
extension $E$.

\begin{eqnarray*}
d_3 &=& \frac{W:A}{A \vee B \vee C} \\
\\
d_4 &=& \frac{A \vee B \vee C:B}{A \vee B}
\end{eqnarray*}

The processes of $W'$ are $[d_1d_3]$ and $[d_2]$. There is
no way to avoid the first default $d_3$ of the new extension
to be part of some process of $W'$ as well. However, as in
this case, it may have no effects. Indeed, the extensions
are $A \vee C$ and $B \vee D$. Let us now consider which
defaults can be applied to $W$. Both processes of $W'$ are
still processes of $W$. However, $d_3$ is applicable to $W$,
leading to a state in which both $d_1$ and $d_2$ are
applicable. However, both processes only contain models that
are among the previous ones.

The above terms can be translated into the following
formuale:

\begin{eqnarray*}
A &=& abc \\
B &=& ab\neg c \\
C &=& a\neg bc \\
D &=& a\neg b\neg c
\end{eqnarray*}

The defaults will be then defined as follows.

\begin{eqnarray*}
d_1 &=& \frac{:\neg b \wedge c}{a \wedge c} \\
\\
d_2 &=& \frac{:\neg b \wedge \neg c}{a \wedge \neg c} \\
\\
d_3 &=& \frac{a:b \wedge c}{b \vee c} \\
\\
d_4 &=& \frac{a \wedge (b \vee c):b \wedge \neg c}{b}
\end{eqnarray*}

In $W'$, only $d_1$ and $d_2$ are applicable. The first one
leads to $a \wedge c$, which is consistent with the
justification of $d_3$. The first selected process of $W'$
is therefore $[d_1d_3]$, leading to the extension $a \wedge
c$.

The second process from $W'$ starts with $d_2$, which
generates $a \wedge \neg c$, which is not consistent with
$d_1$ and $d_3$, and does not imply the precondition of
$d_4$. As a result, $[d_2]$ is the second selected process
of $W'$, leading to the extension $a \wedge \neg c$. We
therefore have $\vee \extd(W') \equiv (a \vee c) \wedge (a
\vee \neg c) \equiv a$.

Let us now consider the extensions from $W$. All selected
processes of $W'$ are also selected processes of $W$.
However, we can now apply $d_3$, as $a$ is true in the
background theory. We therefore obtain $b \vee c$. This
conclusion is inconsistent with the justification of $d_2$,
but $d_1$ and $d_4$ can be applied. The first one leads to
the extension $a \wedge c$, which is also an extension of
$\l D,W' \r$. On the other hand, $[d_3d_4]$ leads to $a
\wedge b$, which is a new extension. Nevertheless, $\vee
\extd(W) \equiv (a \vee c) \wedge (a \vee \neg c) \wedge (a
\vee b) \equiv a$: the theory $\l D,W \r$ has some
extensions that $\l D,W' \r$ does not have, but the
skeptical consequences are the same.~\qed

In the proof, we used two defaults that are applicable in
$W$ but not in the processes of $\l D,W' \r$. These two
defaults cannot have mutually consistent justifications;
otherwise, they would be both applicable in some process of
$\l D,W' \r$ thanks to the fact that any extension of $\l
D,W \r$ contains only models of some extensions of $\l D,W'
\r$. This proof does not work for constrained and rational
default logic; however, the same claim can be proved in a
different way.

\begin{theorem}
\label{consequence-not-faithful-global}

There exists $D$, $W$, and $W' \subset W$ such that
$W' \equiv_D^c W$ but
$W' \not\equiv_D^e W$, for
constrained and rational default logic.

\end{theorem}

\proof The idea is as follows: the constrained extensions of
a default theory are each characterized by a model that is
consistent with all justifications and consequences of the
defaults used to generate the extensions
\cite{libe-extensions}. Therefore, we might have a situation
like the one depicted below:

\

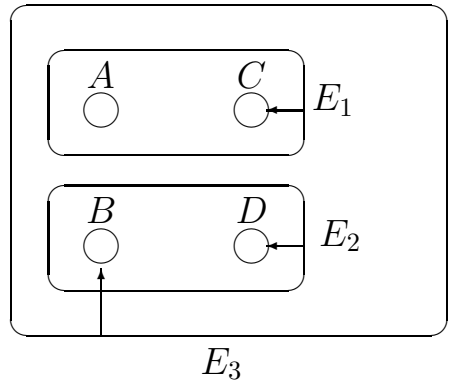
\hfil
%%%%%%%%%%%%% begin:constrained.latex %%%%%%%%%%%%%%
\setlength{\unitlength}{4144sp}%
\begingroup\makeatletter\ifx\SetFigFont\undefined%
\gdef\SetFigFont#1#2#3#4#5{%
  \reset@font\fontsize{#1}{#2pt}%
  \fontfamily{#3}\fontseries{#4}\fontshape{#5}%
  \selectfont}%
\fi\endgroup%
\begin{picture}(2634,2280)(1249,-1969)
{\color[rgb]{0,0,0}\thinlines
\put(1801,-331){\circle{202}}
}%
{\color[rgb]{0,0,0}\put(2701,-331){\circle{202}}
}%
{\color[rgb]{0,0,0}\put(1801,-1141){\circle{202}}
}%
{\color[rgb]{0,0,0}\put(2701,-1141){\circle{202}}
}%
{\color[rgb]{0,0,0}\put(1366,-1576){\oval(210,210)[bl]}
\put(1366,194){\oval(210,210)[tl]}
\put(3766,-1576){\oval(210,210)[br]}
\put(3766,194){\oval(210,210)[tr]}
\put(1366,-1681){\line( 1, 0){2400}}
\put(1366,299){\line( 1, 0){2400}}
\put(1261,-1576){\line( 0, 1){1770}}
\put(3871,-1576){\line( 0, 1){1770}}
}%
{\color[rgb]{0,0,0}\put(1591,-496){\oval(210,210)[bl]}
\put(1591,-76){\oval(210,210)[tl]}
\put(2911,-496){\oval(210,210)[br]}
\put(2911,-76){\oval(210,210)[tr]}
\put(1591,-601){\line( 1, 0){1320}}
\put(1591, 29){\line( 1, 0){1320}}
\put(1486,-496){\line( 0, 1){420}}
\put(3016,-496){\line( 0, 1){420}}
}%
{\color[rgb]{0,0,0}\put(1591,-1306){\oval(210,210)[bl]}
\put(1591,-886){\oval(210,210)[tl]}
\put(2911,-1306){\oval(210,210)[br]}
\put(2911,-886){\oval(210,210)[tr]}
\put(1591,-1411){\line( 1, 0){1320}}
\put(1591,-781){\line( 1, 0){1320}}
\put(1486,-1306){\line( 0, 1){420}}
\put(3016,-1306){\line( 0, 1){420}}
}%
{\color[rgb]{0,0,0}\put(3016,-331){\vector(-1, 0){225}}
}%
{\color[rgb]{0,0,0}\put(3016,-1141){\vector(-1, 0){225}}
}%
{\color[rgb]{0,0,0}\put(1801,-1681){\vector( 0, 1){405}}
}%
\put(1801,-196){\makebox(0,0)[b]{\smash{{\SetFigFont{14}{16.8}{\familydefault}{\mddefault}{\updefault}{\color[rgb]{0,0,0}$A$}%
}}}}
\put(1801,-1006){\makebox(0,0)[b]{\smash{{\SetFigFont{14}{16.8}{\familydefault}{\mddefault}{\updefault}{\color[rgb]{0,0,0}$B$}%
}}}}
\put(2701,-196){\makebox(0,0)[b]{\smash{{\SetFigFont{14}{16.8}{\familydefault}{\mddefault}{\updefault}{\color[rgb]{0,0,0}$C$}%
}}}}
\put(2701,-1006){\makebox(0,0)[b]{\smash{{\SetFigFont{14}{16.8}{\familydefault}{\mddefault}{\updefault}{\color[rgb]{0,0,0}$D$}%
}}}}
\put(2521,-1906){\makebox(0,0)[b]{\smash{{\SetFigFont{14}{16.8}{\familydefault}{\mddefault}{\updefault}{\color[rgb]{0,0,0}$E_3$}%
}}}}
\put(3061,-331){\makebox(0,0)[lb]{\smash{{\SetFigFont{14}{16.8}{\familydefault}{\mddefault}{\updefault}{\color[rgb]{0,0,0}$E_1$}%
}}}}
\put(3106,-1141){\makebox(0,0)[lb]{\smash{{\SetFigFont{14}{16.8}{\familydefault}{\mddefault}{\updefault}{\color[rgb]{0,0,0}$E_2$}%
}}}}
\end{picture}%
%%%%%%%%%%%%% end:constrained.latex %%%%%%%%%%%%%%
 %

\

The arrows indicate the model associated to each extension:
$C$ is the model associated with $E_1$, etc. Note that $E_1
\vee E_2 \vee E_3 \equiv E_1 \vee E_2$. As a result, a
default theory having only the extensions $E_1$ and $E_2$ is
equivalent to a theory having all three extensions, but yet
these two theories have the same consequences.

We define $\l D,W \r$ in such a way it has all three
extensions, but $\l D,\emptyset \r$ does not have the
extension $E_3$ because the model $B$ is excluded by a
default that generates $W$. The above condition can be
realized using two variables $a$ and $b$ to distinguish the
four models $A$, $B$, $C$, and $D$, and a variable $x$ to
distinguish $W$ from $W'$.

\begin{eqnarray*}
D &=& \{d_1, d_2, d_3, d_4\} \hskip 6cm\\
W &=& \{x\} \\
W' &=& \emptyset \\
& \mbox{\hbox to 0pt{the defaults are:}} & \\
\hbox to 0pt{\hskip 5cm generates $W$ from $W'$\hss}
&& d_1 = \frac{:x \wedge a}{x} \\
\hbox to 0pt{\hskip 5cm generates the extension $E_1=Cn(x \wedge b)$\hss}
&& d_2 = \frac{x:a \wedge b}{b}  \\
\hbox to 0pt{\hskip 5cm generates the extension $E_2=Cn(x \wedge \neg b)$\hss}
&& d_3 = \frac{x:a \wedge \neg b}{\neg b} \\
\hbox to 0pt{\hskip 5cm generates the extension $E_3=Cn(x)$\hss}
&& d_4 = \frac{x:\neg a \wedge \neg b}{x}
\end{eqnarray*}

The justification of $d_2$, $d_3$, and $d_4$ are mutually
inconsistent. In $\l D,W \r$, the three extensions are
generated by the processes $[d_2,d_1]$, $[d_3,d_1]$, and
$[d_4]$. The presence of $d_1$ in these processes do not
change their generated extensions, as $\cons(d_1)=x$, which
is already in $W$. We have $E_1 \vee E_2 \vee E_3 \equiv x$.

Let us now consider $\l D,W' \r$. The only default that is
applicable in $W' = \emptyset$ is $d_1$, which generates $x$
but also have $a$ as a justification. As a result, the
defaults $d_2$ and $d_3$ are still applicable, but $d_4$ is
not. As a result, the only extensions of $\l D,W' \r$ are
$E_1$ and $E_2$. We therefore have $W' \not\equiv_D^e W$. On
the other hand, $E_1 \vee E_2 \equiv x$, which is equivalent
to $E_1 \vee E_2 \vee E_3$. As a result, $W' \equiv_D^c
W$.~\qed

\subsubsection{Correspondence, Particular Cases}

While $\equiv_D^c$ and $\equiv_D^e$ are not the same in
general, they coincide when all defaults are normal and one
formula is contained in the other one.

\begin{theorem}
\label{same-constrained}

If $W' \subseteq W$ and $D$ is a set of
normal defaults, then
$W' \equiv_D^c W$ implies
$W' \equiv_D^e W$ in constrained default logic.

\end{theorem}

\proof Given the previous result, we only have to prove that
$W \equiv^c_D W'$ implies that $\extd(W) \subseteq
\extd(W')$, that is, $\l D,W \r$ does not have any extension
that is not an extension of $\l D,W' \r$.

To the contrary, assume that such extension exists. Let
$\Pi$ be the process that generates the extension of $\l D,W
\r$ that is not an extension of $\l D,W' \r$. By definition
of process, $\cons(\Pi) \cup W \cup \just(\Pi)$ is
consistent. Therefore, it has a model $M$. Since this model
satisfies both $W$ and $\cons(\Pi)$, it is a model of the
extension generated by $\Pi$.

Since the conclusions of the two theories are the same,
every model of the extension generated by $\Pi$ is a model
of some extensions of $\l D,W' \r$. Let $\Pi'$ be the a
process of $\l D,W' \r$ that generates an extension that
contains the model $M$. We prove that all defaults of $\Pi$
are in $\Pi'$.

Since $M$ is a model of the extension generated by $\Pi'$,
it is a model of $\cons(\Pi') \cup W'$. Therefore, it is a
model of $\cons(\Pi')$, and a model of $\just(\Pi')$ because
defaults are normal. We have already proved that $M$ is a
model of $\cons(\Pi)$ and $\just(\Pi)$ and of $W$. As a
result, the set $\cons(\Pi) \cup \cons(\Pi') \cup W \cup
just(\Pi) \cup \just(\Pi')$ is consistent. As a result, we
can add all defaults of $\Pi$ to $\Pi'$ without
contradicting the justifications.

As a result, the defaults of $\Pi$ are not in $\Pi'$ only if
their preconditions are not entailed from the consequences
of $\Pi$. This is impossible: since $\Pi$ is a process of
$\l D,W \r$, we have $W \models \prec(d)$, where $d$ is the
first default of $\Pi$. As a result, $d$ must be part of
$\Pi'$, otherwise $\Pi'$ would not be a maximal process. The
consequences of $d$ are therefore part of $\cons(\Pi') \cup
W'$. Repeating the argument with the second default of $\Pi$
we get the same result. We can therefore conclude that all
defaults of $\Pi$ are in $\Pi'$.~\qed

Since Reiter, justified, constrained, and rational default
logics coincide on normal default theories, the equality of
the definitions of equivalence holds when defaults are
normal.

\begin{theorem}
\label{equal-normal}

If $D$ is a set of normal defaults and $W' \subseteq W$,
then $W' \equiv_D^c W$ if and only if $W' \equiv_D^e W$.

\end{theorem}

\proofnewpage

When all defaults are categorical (prerequisite-free), the
following lemma allows proving that the three considered
forms of equivalence coincide.

\begin{lemma}

If $D$ is a set of categorical defaults, $W' \subseteq
W$, and $W' \infdef W$, then $W \infdef^e W'$ in constrained
default logic.

\end{lemma}

\proof Let $\Pi$ be a selected process of $\l D,W \r$. We
prove that $\Pi$ is a selected process of $\l D,W' \r$
generating the same extension.

Since $\Pi$ is a selected process of $\l D,W \r$, it holds
that $W \cup \cons(\Pi) \cup \just(\Pi)$ is consistent.
Since $W' \subseteq W$, it also holds that $W' \cup
\cons(\Pi) \cup \just(\Pi)$ is consistent. Since no default
has preconditions, $\Pi$ is a successful process of $\l D,W'
\r$. Since constrained default logic is a failsafe semantics
\cite{libe-failsafe}, there exists $\Pi'$ such that $\Pi
\cdot \Pi'$ is a selected process of $\l D,W \r$.

Since every extension of $W'$ entails $W$, this is in
particular true for the extension generated by $\Pi \cdot
\Pi'$. In other words, $W' \cup \cons(\Pi \cdot \Pi')
\models W$. As a result, $W' \cup \cons(\Pi \cdot \Pi')
\equiv W \cup \cons(\Pi \cdot \Pi')$. Since $\Pi \cdot \Pi'$
is a process of $\l D,W' \r$, we have that $W' \cup
\cons(\Pi \cdot \Pi') \cup \just(\Pi \cdot \Pi')$ is
consistent, which is therefore equivalent to the consistency
of $W \cup \cons(\Pi \cdot \Pi') \cup \just(\Pi \cdot
\Pi')$. Therefore, $\Pi \cdot \Pi'$ is a successful process
of $\l D,W \r$. Since $\Pi$ is by assumption a maximal
successful process of $\l D,W \r$, it must be $\Pi'=[~]$,
that is, $\Pi \cdot \Pi' = \Pi$. We have already proved that
$W' \cup \cons(\Pi \cdot \Pi') \equiv W \cup \cons(\Pi \cdot
\Pi')$, that is, $\Pi$ generates the same extension in $W$
and in $W'$.~\qed

Since constrained and Reiter default logics coincide on
normal default theories, we have the following consequence.

\begin{corollary}
\label{all-the-same}

If $D$ is a set of normal and categorical defaults and $W'
\subseteq W$, the conditions $W' \equiv_D^m W$, $W'
\equiv_D^c W$, and $W' \equiv_D^e W$ are equivalent.

\end{corollary}

%%%%%%%%%%%%% end:default-equiv %%%%%%%%%%%%%%
 %

%%%%%%%%%%%%% begin:default-one %%%%%%%%%%%%%%
\subsection{Redundancy of Clauses vs.\  Theories}

The redundancy of a clause $\gamma$ in a formula $W$ is
defined as the equivalence of $W$ and $W \backslash
\{\gamma\}$. The redundancy of a formula $W$ is defined as
its equivalence to one of its proper subsets. A formula
containing a redundant clause is redundant, but the converse
is not always true: a formula might contain no redundant
clause but yet it is equivalent to one of its proper
subsets.

In this section, we compare the redundancy of a set of
clauses with the redundancy of a single clause in default
logic. In propositional logics, these two concepts are the
same: $\Pi$ is equivalent to one of its proper subsets if
and only if it contains a redundant clause. In default
logic, it may be that $\gamma_1$ and $\gamma_2$ are both
irredundant in $\{\gamma_1,\gamma_2\}$ while
$\{\gamma_1,\gamma_2\}$ is redundant, as shown by the theory
$\l D, W \r$ defined below.

\begin{eqnarray*}
W &=& \{a,b\} \\
D &=& \{d_1,d_2,d_3\} \\
& \hbox to 0pt{where} \\
&& d_1 = \frac{a:\neg b}{\neg b} \\
&& d_2 = \frac{b: \neg a}{\neg a} \\
&& d_3 = \frac{:a \wedge b}{a \wedge b}
\end{eqnarray*}

The theory $\l D, W \r$ has the single extension
$Cn(\{a,b\})$. Indeed, $d_1$ and $d_2$ are not applicable
because their justifications are inconsistent with $W$. The
third default is applicable, but its consequence is $a
\wedge b$, which is already in the theory.

The theory $\l D,W \backslash \{b\} \r$ still has the
extension $\{a,b\}$, which results from the application of
$d_3$, which then blocks the application of $d_1$ and $d_2$.
However, it also has a new extension: since $d_1$ is
applicable, it generates $\neg b$, which blocks the
application of $d_3$. This produces the extension $\{a,\neg
b\}$. In the same way, $\l D, W \backslash \{a\} \r$ has the
two extensions $\{a,b\}$ and $\{\neg a,b\}$

The theory $\l D, W \backslash \{a,b\} \r$ has again a
single extension: $d_3$ is the only applicable default,
leading to the addition of $a \wedge b$. Neither $d_1$ nor
$d_2$ are applicable. Therefore, $\{a,b\}$ is the only
extension of this theory.

The set of extensions of the theory is changed by removing
any single clause, but is not changed by the removal of both
clauses. In other words, both $a$ and $b$ are irredundant in
$\{a,b\}$, but $\{a,b\}$ is redundant. Since $D$ is a set of
normal default, this counterexample holds even for normal
default theories.

\proofnewpage

The two theories obtained by removing a single clause of $W$
differ from $W$ because of a new extension. This can be
proved to be always the case if the removal of both clauses
leads to the original set of extensions. This is proved by
first showing a sort of ``continuity'' of extensions.

\begin{lemma}
\label{process-monotone}

If $E$ is an extension of $\l D,W' \r$ and of $\l D,W \r$
with $W' \subseteq W$, then every selected process of $\l
D,W' \r$ generating $E$ is a selected process of $\l D,W \r$
generating $E$.

\end{lemma}

\proof Let $\Pi$ be a selected process of $\l D,W' \r$ that
generates $E$. Since $W' \subseteq W$ we have $W \models
W'$; therefore, $\Pi$ is a process of $\l D,W \r$. Remains
to prove that it is also selected. However, all conditions
for a process to be selected in $\l D,W' \r$ contains $W'$
only in the subformula $W' \cup \cons(\Pi)$. Since $E = W'
\cup \cons(\Pi)$, and $E$ is an extension of $\l D,W \r$, we
have that $E \models W$. As a result, $W' \cup \cons(\Pi)
\equiv W \cup \cons(\Pi)$. Therefore, every condition for
$\Pi$ in $\l D,W' \r$ is equivalent to the same condition
for $\l D,W \r$.~\qed

The following lemma relates the selected processes of three
formulae.

\begin{lemma}
\label{process-median}

If $\Pi$ is a selected process of both $\l D,W' \r$ and $\l
D,W \r$ and generates the same extension in both theories,
it is also a selected process of every $\l D,W'' \r$ with
$W' \subseteq W'' \subseteq W$ and generates the same
extension in $\l D,W'' \r$.

\end{lemma}

\proof If $W' \subseteq W$ does not hold, the claim is
trivially true because there is no $W''$ such that $W'
\subseteq W'' \subseteq W$.

Since $\Pi$ is a process of $\l D,W' \r$, it is also a
process of $\l D,W'' \r$ because $W'' \models W'$. Since
$\Pi$ generates the same extensions in $W'$ and $W$, we have
that $W' \cup \cons(\Pi) \equiv W \cup \cons(\Pi)$. Since $W
\models W'' \models W'$, we also have $W' \cup \cons(\Pi)
\equiv W'' \cup \cons(\Pi)$. Therefore, every condition that
is true for $W' \cup \cons(\Pi)$ is also true for $W'' \cup
\cons(\Pi)$.~\qed

The following lemma proves that extensions of both a theory
and a subset of it are also extensions of any theory
``between them''.

\begin{lemma}
\label{extension-continue}

If $E$ is an extension of both $\l D,W'\r$
and $\l D,W \r$, with $W' \subseteq W$,
then it is an extension of any
$\l D,W'' \r$ with $W' \subseteq W'' \subseteq W$

\end{lemma}

\proof By Lemma~\ref{process-monotone}, every selected
process $\Pi$ of $\l D,W' \r$ that generates $E$ is also a
selected process of $\l D,W \r$ and generates the same
extension $E$ in this theory. As a result,
Lemma~\ref{process-median} applies, and $\Pi$ is a selected
process of $\l D,W'' \r$ and generates the same
extension.~\qed

This theorem shows that extensions have a form of ``partial
monotonicity'': an extension of both a subset and a superset
of a formula is also an extension of the formula. This is
important to our aims, as it shows that the equivalence $W'
\equiv_D^e W$ implies that all default theories $\l D,W''
\r$ with $W' \subseteq W'' \subseteq W$ have the same
extensions of $\l D,W \r$. Therefore, $\l D,W'' \r$ can
differ from $\l D,W \r$ only because of new extensions.

\begin{corollary}

If $W' \equiv_D^e W$, $W' \subseteq W'' \subseteq W$, and
$W'' \not\equiv_D^e W$, then $\extd(W) \subset \extd(W'')$.

\end{corollary}

The existence of extensions of $W''$ that are not extensions
of $W$ does not imply that $W''$ theory is not
consequence-equivalent to $W$ and $W''$. On the other hand,
$W'' \not\equiv_D^c W$ implies $W'' \not\equiv_D^e W$, which
leads to the following consequence.

\begin{corollary}

If $W' \equiv_D^e W$, $W' \subseteq W'' \subseteq W$, and
$W'' \not\equiv_D^c W$, then $\extd(W) \subset \extd(W'')$.

\end{corollary}

While it is not true that the irredundancy of two clauses
proves the irredundancy of the set composed of them, it is
however true that this can only happen because of new
extensions that are created by removing a single clause. For
some special cases of default logics, such creation is not
possible.

\begin{theorem}

If $D$ is a set of normal and categorical defaults, then $W'
\equiv_D^e W$ implies that $W \equiv_D^e W''$ for any $W''$
such that $W' \subseteq W'' \subseteq W$.

\end{theorem}

\proof By Lemma~\ref{extension-continue}, all extensions of
$\l D,W \r$ are also extensions of $\l D,W'' \r$. We
therefore only have to prove the converse. Let $\Pi$ be a
process of $\l D,W'' \r$. Since the theory has no
preconditions, all defaults $d \in \Pi$ satisfy $W' \cup
\cons(\Pi[d]) \models d$. In other words, $\Pi$ is a process
of $\l D,W' \r$. Since $W'' \cup \cons(\Pi) \cup \just(d)$
is consistent for every $d \in \Pi$, and $W'$ is logically
weaker than $W''$, the same condition is true for $W'$.
Since normal default logic is fail-safe
\cite{libe-failsafe}, there exists $\Pi'$ such that $\Pi
\cdot \Pi'$ is a selected process of $\l D,W' \r$. By
Lemma~\ref{process-monotone} and
Lemma~\ref{extension-continue}, $W \cup \cons(\Pi \cdot
\Pi')$ is an extension of $\l D,W'' \r$.

Let us assume that $W \cup \cons(\Pi \cdot \Pi')$ and $W''
\cup \cons(\Pi)$ are not equivalent. This is only possible
if $W \cup \cons(\Pi \cdot \Pi') \models W'' \cup
\cons(\Pi)$ but not vice versa. This is however impossible,
because in normal default logic all extensions are mutually
inconsistent \cite{reit-80}.~\qed

Since the two definitions of equivalence are the same on
normal default theories, as proved in
Theorem~\ref{equal-normal}, this result extends to the
definition of redundancy based on consequences.

\begin{corollary}

If $D$ is a set of normal and categorical defaults, then
$W \equiv_D^c W''$ implies that
$W \equiv_D^c W'$ for any $W'$
such that $W'' \subseteq W' \subseteq W$.

\end{corollary}

This result does not hold for normal default theories with
preconditions, as the counterexample at the beginning of the
section is only composed of normal defaults with
preconditions.

\begin{corollary}

If $D$ is a set of normal and categorical defaults, then a
formula is redundant if and only if it contains a
redundant clause.

\end{corollary}

In other words, default logic restricted to the case of
normal and categorical defaults has the local redundancy
property.

%%%%%%%%%%%%% end:default-one %%%%%%%%%%%%%%
 %

%%%%%%%%%%%%% begin:default-irred %%%%%%%%%%%%%%
\subsection{Making Clauses Irredundant}

Modifying a theory in order to make some parts irredundant
proved useful for classical and circumscriptive logics. We
show a similar result for default logic.

\begin{definition}

The $M$-irredundant version of a default theory $\l D,W \r$,
where $M \subseteq W = \{\gamma_1,\ldots,\gamma_m\}$, is the
following theory, where $\{c_1,\ldots,c_m\}$ are new
variables.

\begin{eqnarray*}
I(\l D,W \r,M) = \l D',W' \r \\
\mbox{where:\hss} \\
W' &=&
\{c_i \vee \gamma_i ~|~ \gamma_i \in W\}
\\
D' &=& D_1 \cup D_2 \cup D_3
\\
D_1 &=&
\left\{
\frac{:\neg c_1 \wedge \cdots \wedge \neg c_m}
{\neg c_1 \wedge \cdots \wedge \neg c_m}
\right\} 
\cup
\\
D_2 &=& 
\left\{
\left.
\frac
{c_i \vee \gamma_i:c_i \wedge \{\neg c_i ~|~ 1 \leq j \leq k ,~ i \not= i\}}
{c_i \wedge \{\neg c_i ~|~ 1 \leq j \leq k ,~ i \not= i\}}
\right|
\gamma_i \in M
\right\} \cup \\
D_3 &=&
\left\{
\left.
\frac{\neg c_1 \wedge \cdots \wedge \neg c_m \wedge \alpha:\beta}{\gamma} 
\right|
\frac{\alpha:\beta}{\gamma} \in D
\right\}
\end{eqnarray*}

\end{definition}

The clauses of $M$ are made irredundant by this
transformation, while the redundancy of the other clauses
does not change.

\begin{theorem}
\label{default-irredundant}

If $M \subseteq W$, $\l D',W' \r=I(\l D,W \r,M)$, and $W''
\subset W$, then $W' \equiv_{D'}^e W''$ holds if and only if
$\{\gamma_i ~|~ c_i \vee \gamma_i \in W''\} \equiv_D^e W$
and $W''$ contains all clauses $c_i \vee \gamma_i$ such that
$\gamma_i \in M$. The same holds for
consequence-equivalence.

\end{theorem}

\proof The default of $D_1$ can be applied provided that $W$
is consistent. Its application makes all defaults of $D_2$
inapplicable and makes the background theory and the
defaults of $D_3$ equivalent to $W$ and to $D$,
respectively. As a result, $I(\l D,W \r,M)$ has an extension
$Cn(E \wedge \neg c_1 \wedge \cdots \wedge \neg c_m)$ for
any extension $E$ of $\l D,W \r$. As a result, a subset of
$W'$ has this extension if and only if the corresponding
subset of $W$ has the same extension.

The $i$-th default of $D_2$ is applicable to $W$ because
$c_i \vee \gamma_i$ is in the background theory. Since all
clauses of $W$ contain the literals $c_j$ only positively,
these literals cannot be removed by resolution. As a result,
every non-tautological consequence of $W \backslash \{c_i
\vee \gamma_i\}$ is disjoined with at least a variable $c_j$
with $j \not= i$. As a result, no subset of $W'$ allows for
the application of this default unless it contains the
clause $c_i \vee \gamma_i$.

The application of this default makes all other defaults of
$D'$ inapplicable. The generated extension is moreover
inconsistent with all other extensions of the theory. As a
result, any subset of $W'$ not containing $c_i \vee
\gamma_i$ necessarily has a different set of extensions and
consequences than $W'$.~\qed

\comment

Not used because the proof of hardness of clause redundancy
use a single clause.

\endcomment

%%%%%%%%%%%%% end:default-irred %%%%%%%%%%%%%%
 %

%%%%%%%%%%%%% begin:default-clause %%%%%%%%%%%%%%
\subsection{Complexity of Clause Redundancy}

In this section, we analyze the complexity of checking the
redundancy of a clause in a formula. Formally, this is the
problem of whether $W \backslash \{\gamma\}$ is equivalent
to $W$ according to $\equiv_D^c$ or $\equiv_D^e$. By
Corollary~\ref{entail-same}, these two forms of equivalence
are related, as $W' \infdef^c W$ is equivalent to $W'
\infdef^e W$ and also to $W' \infdef W$, if $W' \subseteq
W$. As a result, checking whether $W' \infdef W$ allows for
telling whether the ``first part of equivalence'' between
$W'$ and $W$ holds, for both kinds of equivalence. In other
words, in order to check whether $W'$ and $W$ are equivalent
with $W' \subseteq W$, we can first check whether $W'
\infdef W$; if this condition is true, we then proceed
checking whether $W \infdef^c W'$ or $W \infdef^e W'$
depending on which equivalence is considered.

Lemma~\ref{derive-theory} tells that $W' \infdef W$ implies
that all processes of $\l D,W' \r$ are also processes of $\l
D,W \r$. This condition does not imply equivalence because
$\l D,W \r$ may contain some other processes, as in the
default theory $\l D,W \r$ below.

\begin{eqnarray*}
W &=& \{a\} \\
D &=& \{d_1,d_2\} \\
& \hbox to 0pt{where:\hss} \\
&& d_1 = \frac{a:b}{b} \\
&& d_2 = \frac{:a \wedge \neg b}{a \wedge \neg b}
\end{eqnarray*}

The theory $\l D,W \r$ has two extensions: applying either
$d_1$ or $d_2$, the other is not applicable. The resulting
extensions are $Cn(a \wedge b)$ and $Cn(a \wedge \neg b)$.
Let $W' = \emptyset$. The only default that is applicable in
$W'$ is $d_2$, leading to the only extension $Cn(a \wedge
\neg b)$. This extension implies $W$. As a result, we have
that $W' \infdef W$ but $W'$ and $W$ do not have the same
extensions and the same consequences. In particular, $W$ has
some extensions that $W'$ does not have. This is always the
case if $W' \infdef W$ but $W$ and $W'$ are not equivalent.

In order to check equivalence of $W'$ and $W$ with $W'
\subseteq W$, two conditions have to be checked:

\begin{enumerate}

\item $W' \infdef W$; and

\item $W \infdef^c W'$ or $W \infdef^e W'$.

\end{enumerate}

An upper bound on the complexity of checking the redundancy
of a clause is given by the following theorem.

\begin{theorem}
\label{equiv-exte-membership}

Checking whether $W' \equiv_D^e W$ in Reiter and justified
default logic is in \P{2}\  if $W' \subseteq W$.

\end{theorem}

\proof Checking whether $W' \infdef W$ is in \P{2}. The
other condition to be checked is $W \models_D^e W'$. The
converse of this condition is that there exists a formula $E
\subseteq W \cup \cons(D)$ such that $E$ is an extension of
$\l D,W \r$ but is not an extension of $\l D,W ' \r$. Since
checking whether a formula is a Reiter or justified default
logic is in \Dlog{2} \cite{rosa-99}, the whole problem is in
\S{2}. Its converse is therefore in \P{2}. The problem of
redundancy of a clause can be solved by solving two problems
in \P{2}\  in parallel.~\qed

The hardness of the problem for the same class is proved by
the following theorem.

\begin{theorem}
\label{one-hard-exte}

Checking whether $W' \equiv_D^e W$ is \P{2}-hard even if $W
= W' \cup \{a\}$ and all defaults are categorical and
normal.

\end{theorem}

\proof The claim could be proved from the fact that
entailment in default logic is \P{2}-hard even if the
formula to entail is a single positive literal, and all
defaults are categorical and normal
\cite{gott-92-b,stil-92-b}. If all defaults are categorical
and normal, Corollary~\ref{all-the-same} proves that $W'
\equiv_D^m W$ is equivalent to the two other forms of
equivalence.

\iffalse

We show a reduction from $W' \infdef a$ into the problem of
checking the redundancy of the clause $a$ in the formula $W
= W' \cup \{a\}$. Since $W' \subset W$ and all extensions of
$\l D,W \r$ entail $W$, they entail $W'$ as well. In other
words, $W \infdef W'$. If $W' \infdef a$, we have that $W'
\infdef W' \cup \{a\}$, and therefore $a$ is redundant
according to all definitions of equivalence. If $W'
\not\infdef a$, then $W' \not\infdef W' \cup \{a\}$ as well.
Therefore, $a$ is not redundant according to all forms of
equivalence.

\fi

We however use a new reduction from $\forall\exists$QBF
because this is required by the proof of \S{3}-hardness of
formula redundancy. The formula $\forall X \exists Y ~.~ F$
is valid if and only if $a$ is redundant in the theory
below:

\[
\left\l
\left\{
\frac{:x_i}{x_i}, \frac{:\neg x_i}{\neg x_i}
\right\} 
\cup
\left\{
\frac{:F \wedge a}{a}
\right\}
,~
\{a\}
\right\r
\]

This theory has an extension for every possible truth
evaluation over the variables $X$. For each such extension,
the last default can be applied only if $F$ is consistent
with the given evaluation of $X$. As a result, if $F$ is
consistent with every truth evaluation over the variables
$X$, then $a$ can be removed from the background theory
without changing the consequences of these extensions.
Otherwise, the removal of $a$ would cause some of these
extensions not to entail $a$ any more.~\qed

We now consider the problem of redundancy of clauses when
consequence-equivalence is used. The difference between
the two kinds of equivalence is that two sets of
extensions may be different but yet their disjunctions are
the same. The necessity of calculating the disjunction of
all extensions intuitively explains why checking redundancy
for consequence-equivalence is harder than for faithful
equivalence.

\begin{theorem}

The problem of checking whether $W' \equiv_D^c W$ is in
\P{3}\  if $W' \subseteq W$.

\end{theorem}

\proof $W'$ and $W$ are consequence-equivalent if $W'
\infdef W$ and $W \models_D^c W'$. The first problem is in
\P{2}. We prove that the converse of the second condition is
in \S{3}. By definition, $W \not\models_D^c W'$ holds if and
only if $\vee \extd(W) \not\models \vee \extd(W')$. In terms
of models, we have $\cup \{\mod(E) ~|~ E \in \extd(W) \}
\not\subseteq \cup \{\mod(E) ~|~ E \in \extd(W') \}$, that
is, there exists $M$ and $E$ such that $M \in \mod(E)$, $E
\in \extd(W)$, but $M$ is not a model of any extension of
$W'$. The whole condition can therefore be expressed by the
following formula.

\[
\exists M \exists E ~.~
M \in \mod(E) \wedge E \in \extd(W) \wedge
\big(
\forall E' ~.~ E' \not\in \extd(W') \vee M \not\in \mod(E')
\big)
\]

Since $E' \not\in \extd(W')$ is in \Dlog{2}\  for Reiter
\cite{rosa-99} and justified default logic and in \P{2}\ for
constrained and rational \cite{libe-exte}, the problem of
checking $W \infdef^c W'$ is in \P{3}. Therefore, the
problem of consequence-equivalence is in \P{3}\ as well for
all four considered semantics.~\qed 

We show that the problem is hard for the same class.

\begin{theorem}
\label{one-hard-cons}

The problem of checking whether $W' \equiv_D^c W$ for Reiter
and justified default logics is \P{3}-hard even if $W = W'
\cup \{a\}$.

\end{theorem}

\proof Since checking whether $W' \equiv_D^e W$ is in \P{2},
a proof of \P{3}-hardness necessarily requires the use of
theories having different extensions but might have the same
consequences.

We prove that the problem of non-equivalence of default
theories is \S{3}-hard by reduction from QBF. We reduce a
formula $\exists X \forall Y \exists Z . F$ into the problem
of checking whether $W \equiv_D^c W'$, where $W'=\emptyset$,
$W=\{a\}$, and $D=D_1 \cup D_2 \cup D_3 \cup D_4 \cup D_5
\cup D_6$. We show each $D_i$ at time. First, we generate a
complete evaluation over the variables $X$ using the
following defaults.

\begin{eqnarray*}
D_1 &=&
\left\{
\frac{:x_i}{x_i \wedge h_i} ~,~~
\frac{:\neg x_i}{\neg x_i \wedge h_i}
\right\}
\end{eqnarray*}

Since these defaults have no preconditions, they can be
applied regardless of whether $W$ or $W'$ is the background
theory. They generate a process for any truth evaluation
$\omega_X$ over the variables in $X$. The variables $h_i$
are all true only when all variables $x_i$ have been set to
a value.

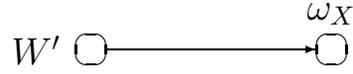
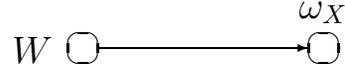
\begin{center}
%%%%%%%%%%%%% begin:cons-redu-1.latex %%%%%%%%%%%%%%
\setlength{\unitlength}{4144sp}%
\begingroup\makeatletter\ifx\SetFigFont\undefined%
\gdef\SetFigFont#1#2#3#4#5{%
  \reset@font\fontsize{#1}{#2pt}%
  \fontfamily{#3}\fontseries{#4}\fontshape{#5}%
  \selectfont}%
\fi\endgroup%
\begin{picture}(2180,1932)(169,-1555)
{\color[rgb]{0,0,0}\thinlines
\put(811, 29){\oval(180,180)}
}%
{\color[rgb]{0,0,0}\put(811,-1411){\oval(180,180)}
}%
{\color[rgb]{0,0,0}\put(2251,-1411){\oval(180,180)}
}%
{\color[rgb]{0,0,0}\put(2251, 29){\oval(180,180)}
}%
{\color[rgb]{0,0,0}\put(901, 29){\vector( 1, 0){1260}}
}%
{\color[rgb]{0,0,0}\put(901,-1411){\vector( 1, 0){1260}}
}%
\put(631,-1501){\makebox(0,0)[rb]{\smash{{\SetFigFont{14}{16.8}{\familydefault}{\mddefault}{\updefault}{\color[rgb]{0,0,0}$W$}%
}}}}
\put(2251,209){\makebox(0,0)[b]{\smash{{\SetFigFont{14}{16.8}{\familydefault}{\mddefault}{\updefault}{\color[rgb]{0,0,0}$\omega_X$}%
}}}}
\put(2251,-1231){\makebox(0,0)[b]{\smash{{\SetFigFont{14}{16.8}{\familydefault}{\mddefault}{\updefault}{\color[rgb]{0,0,0}$\omega_X$}%
}}}}
\put(631,-61){\makebox(0,0)[rb]{\smash{{\SetFigFont{14}{16.8}{\familydefault}{\mddefault}{\updefault}{\color[rgb]{0,0,0}$W'$}%
}}}}
\end{picture}%
%%%%%%%%%%%%% end:cons-redu-1.latex %%%%%%%%%%%%%%
 %

\end{center}

The processes of $W$ and $W'$ are so far the same. Once all
$h_i$ are true, we can apply the defaults of $D_2 =
\{d_1,d_2,d_3,d_4\}$, which are the ones used in
Theorem~\ref{consequence-not-faithful} to show two theories
that have the same consequences but different extensions:

\begin{eqnarray*}
D_2 &=& \{d_1,d_2,d_3,d_4\} \\
\\
d_1 &=& \frac{h_1 \wedge \cdots \wedge h_n:\neg b \wedge c}{a \wedge c} \\
\\
d_2 &=& \frac{h_1 \wedge \cdots \wedge h_n:\neg b \wedge \neg c}{a \wedge \neg c} \\
\\
d_3 &=& \frac{h_1 \wedge \cdots \wedge h_n \wedge a:b \wedge c}{b \vee c} \\
\\
d_4 &=& \frac{h_1 \wedge \cdots \wedge h_n \wedge a \wedge (b \vee c):b \wedge \neg c}{b}
\end{eqnarray*}

Since these default have $h_1 \wedge \cdots \wedge h_n$ as a
precondition, they can only be applied once a truth
assignment over $X$ has been generated by the previous
defaults. They act as in the proof of
Theorem~\ref{consequence-not-faithful}. Only $[d_1d_3]$ and
$[d_2]$ are processes of $W'$; their consequences are $a
\wedge c$ and $a \wedge \neg c$. The theory $W$ has the same
processes, but also $[d_3d_1]$ and $[d_3d_4]$, which
generate the extensions $a \wedge c$ and $a \wedge b$,
respectively. While the first is also an extension of $W'$,
the second is not. The disjunction of all
extensions is equivalent to $a$ for both $W$ and $W'$.

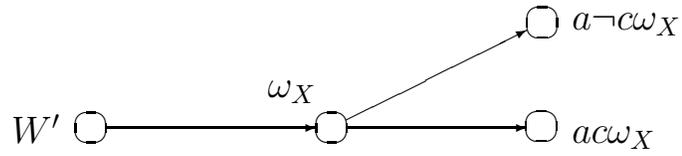
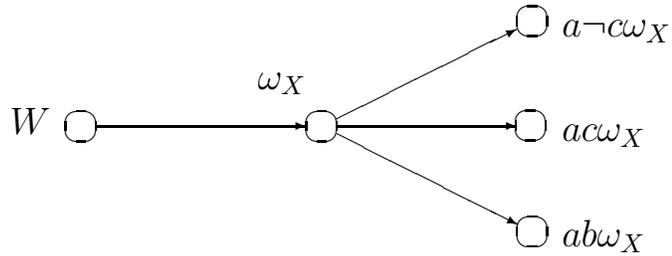
\begin{center}
%%%%%%%%%%%%% begin:cons-redu-2.latex %%%%%%%%%%%%%%
\setlength{\unitlength}{4144sp}%
\begingroup\makeatletter\ifx\SetFigFont\undefined%
\gdef\SetFigFont#1#2#3#4#5{%
  \reset@font\fontsize{#1}{#2pt}%
  \fontfamily{#3}\fontseries{#4}\fontshape{#5}%
  \selectfont}%
\fi\endgroup%
\begin{picture}(3522,3606)(169,-2824)
{\color[rgb]{0,0,0}\thinlines
\put(811, 29){\oval(180,180)}
}%
{\color[rgb]{0,0,0}\put(2251, 29){\oval(180,180)}
}%
{\color[rgb]{0,0,0}\put(3511,659){\oval(180,180)}
}%
{\color[rgb]{0,0,0}\put(811,-2041){\oval(180,180)}
}%
{\color[rgb]{0,0,0}\put(2251,-2041){\oval(180,180)}
}%
{\color[rgb]{0,0,0}\put(3501,-2036){\oval(180,180)}
}%
{\color[rgb]{0,0,0}\put(3511,-1411){\oval(180,180)}
}%
{\color[rgb]{0,0,0}\put(3511,-2671){\oval(180,180)}
}%
{\color[rgb]{0,0,0}\put(3506, 39){\oval(180,180)}
}%
{\color[rgb]{0,0,0}\put(901, 29){\vector( 1, 0){1260}}
}%
{\color[rgb]{0,0,0}\put(2341, 74){\vector( 2, 1){1080}}
}%
{\color[rgb]{0,0,0}\put(901,-2041){\vector( 1, 0){1260}}
}%
{\color[rgb]{0,0,0}\put(2341,-2041){\vector( 1, 0){1080}}
}%
{\color[rgb]{0,0,0}\put(2341,-1996){\vector( 2, 1){1080}}
}%
{\color[rgb]{0,0,0}\put(2345,-2088){\vector( 2,-1){1080}}
}%
{\color[rgb]{0,0,0}\put(2341, 29){\vector( 1, 0){1080}}
}%
\put(631,-61){\makebox(0,0)[rb]{\smash{{\SetFigFont{14}{16.8}{\familydefault}{\mddefault}{\updefault}{\color[rgb]{0,0,0}$W'$}%
}}}}
\put(631,-2086){\makebox(0,0)[rb]{\smash{{\SetFigFont{14}{16.8}{\familydefault}{\mddefault}{\updefault}{\color[rgb]{0,0,0}$W$}%
}}}}
\put(2161,-1816){\makebox(0,0)[rb]{\smash{{\SetFigFont{14}{16.8}{\familydefault}{\mddefault}{\updefault}{\color[rgb]{0,0,0}$\omega_X$}%
}}}}
\put(2161,209){\makebox(0,0)[rb]{\smash{{\SetFigFont{14}{16.8}{\familydefault}{\mddefault}{\updefault}{\color[rgb]{0,0,0}$\omega_X$}%
}}}}
\put(3691,614){\makebox(0,0)[lb]{\smash{{\SetFigFont{14}{16.8}{\familydefault}{\mddefault}{\updefault}{\color[rgb]{0,0,0}$a \neg c\omega_X$}%
}}}}
\put(3691,-1501){\makebox(0,0)[lb]{\smash{{\SetFigFont{14}{16.8}{\familydefault}{\mddefault}{\updefault}{\color[rgb]{0,0,0}$a \neg c\omega_X$}%
}}}}
\put(3691,-2131){\makebox(0,0)[lb]{\smash{{\SetFigFont{14}{16.8}{\familydefault}{\mddefault}{\updefault}{\color[rgb]{0,0,0}$ac\omega_X$}%
}}}}
\put(3691,-2761){\makebox(0,0)[lb]{\smash{{\SetFigFont{14}{16.8}{\familydefault}{\mddefault}{\updefault}{\color[rgb]{0,0,0}$ab\omega_X$}%
}}}}
\put(3691,-61){\makebox(0,0)[lb]{\smash{{\SetFigFont{14}{16.8}{\familydefault}{\mddefault}{\updefault}{\color[rgb]{0,0,0}$ac\omega_X$}%
}}}}
\end{picture}%
%%%%%%%%%%%%% end:cons-redu-2.latex %%%%%%%%%%%%%%
 %

\end{center}

The idea is as follows: from $ab\omega_X$, which is obtained
from $W$ but not from $W'$, we always generate the extension
$ab \omega_X d \epsilon_Y$, where $\epsilon_Y$ is the
assignment of $\false$ to all variables $Y$; from the two
other points $a\neg c\omega_X$ and $ac\omega_X$ we instead
generate an arbitrary assignment $\omega_Y$, which then has
$ab \omega_X d \epsilon_Y$ as a model only if $F$ is
satisfiable.

This way, if there exists a value $\omega_X$ such that for
all $\omega_Y$ the formula $F$ is satisfiable, then there is
no extension of $W'$ having the model
$abd\omega_X\epsilon_Y$. Vice versa, if there exists even a
single $\omega_Y$ such that $F$ is unsatisfiable, an
extension $ac\omega_X\ldots$ for $W'$ will be generated, and
this extension has the model $ab \omega_X d \epsilon_Y$.

\

\noindent%%%%%%%%%%%%% begin:cons-redu-3.latex %%%%%%%%%%%%%%
\setlength{\unitlength}{4144sp}%
\begingroup\makeatletter\ifx\SetFigFont\undefined%
\gdef\SetFigFont#1#2#3#4#5{%
  \reset@font\fontsize{#1}{#2pt}%
  \fontfamily{#3}\fontseries{#4}\fontshape{#5}%
  \selectfont}%
\fi\endgroup%
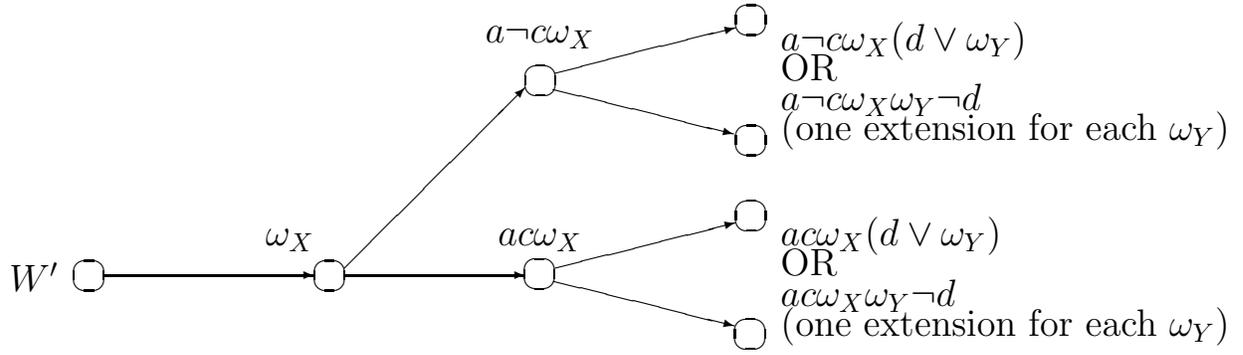
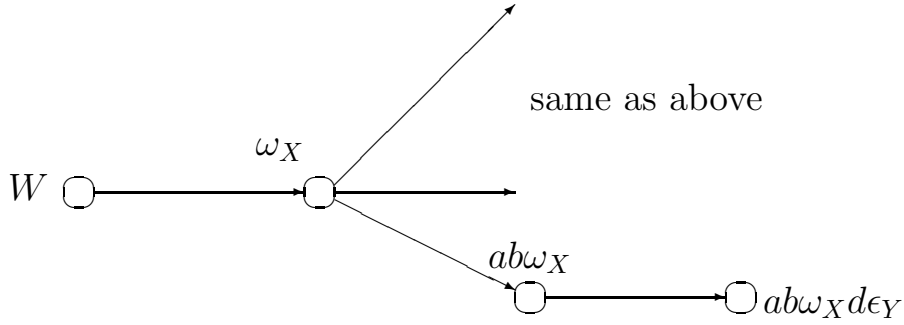
\begin{picture}(5074,4480)(169,-2824)
{\color[rgb]{0,0,0}\thinlines
\put(811, 29){\oval(180,180)}
}%
{\color[rgb]{0,0,0}\put(2251, 29){\oval(180,180)}
}%
{\color[rgb]{0,0,0}\put(811,-2041){\oval(180,180)}
}%
{\color[rgb]{0,0,0}\put(2251,-2041){\oval(180,180)}
}%
{\color[rgb]{0,0,0}\put(3511,-2671){\oval(180,180)}
}%
{\color[rgb]{0,0,0}\put(3506, 39){\oval(180,180)}
}%
{\color[rgb]{0,0,0}\put(4771,389){\oval(180,180)}
}%
{\color[rgb]{0,0,0}\put(4766,-321){\oval(180,180)}
}%
{\color[rgb]{0,0,0}\put(4771,-2671){\oval(180,180)}
}%
{\color[rgb]{0,0,0}\put(3511,1199){\oval(180,180)}
}%
{\color[rgb]{0,0,0}\put(4771,839){\oval(180,180)}
}%
{\color[rgb]{0,0,0}\put(4771,1559){\oval(180,180)}
}%
{\color[rgb]{0,0,0}\put(901, 29){\vector( 1, 0){1260}}
}%
{\color[rgb]{0,0,0}\put(901,-2041){\vector( 1, 0){1260}}
}%
{\color[rgb]{0,0,0}\put(2341,-2041){\vector( 1, 0){1080}}
}%
{\color[rgb]{0,0,0}\put(2345,-2088){\vector( 2,-1){1080}}
}%
{\color[rgb]{0,0,0}\put(2341, 29){\vector( 1, 0){1080}}
}%
{\color[rgb]{0,0,0}\put(3601, 74){\vector( 4, 1){1080}}
}%
{\color[rgb]{0,0,0}\put(3593, -3){\vector( 4,-1){1080}}
}%
{\color[rgb]{0,0,0}\put(3601,-2671){\vector( 1, 0){1080}}
}%
{\color[rgb]{0,0,0}\put(2341, 74){\vector( 1, 1){1080}}
}%
{\color[rgb]{0,0,0}\put(3601,1244){\vector( 4, 1){1080}}
}%
{\color[rgb]{0,0,0}\put(3588,1146){\vector( 4,-1){1080}}
}%
{\color[rgb]{0,0,0}\put(2341,-1996){\vector( 1, 1){1080}}
}%
\put(631,-61){\makebox(0,0)[rb]{\smash{{\SetFigFont{14}{16.8}{\familydefault}{\mddefault}{\updefault}{\color[rgb]{0,0,0}$W'$}%
}}}}
\put(631,-2086){\makebox(0,0)[rb]{\smash{{\SetFigFont{14}{16.8}{\familydefault}{\mddefault}{\updefault}{\color[rgb]{0,0,0}$W$}%
}}}}
\put(2161,-1816){\makebox(0,0)[rb]{\smash{{\SetFigFont{14}{16.8}{\familydefault}{\mddefault}{\updefault}{\color[rgb]{0,0,0}$\omega_X$}%
}}}}
\put(2161,209){\makebox(0,0)[rb]{\smash{{\SetFigFont{14}{16.8}{\familydefault}{\mddefault}{\updefault}{\color[rgb]{0,0,0}$\omega_X$}%
}}}}
\put(3511,209){\makebox(0,0)[b]{\smash{{\SetFigFont{14}{16.8}{\familydefault}{\mddefault}{\updefault}{\color[rgb]{0,0,0}$ac\omega_X$}%
}}}}
\put(3511,-2491){\makebox(0,0)[b]{\smash{{\SetFigFont{14}{16.8}{\familydefault}{\mddefault}{\updefault}{\color[rgb]{0,0,0}$ab\omega_X$}%
}}}}
\put(4906,-2761){\makebox(0,0)[lb]{\smash{{\SetFigFont{14}{16.8}{\familydefault}{\mddefault}{\updefault}{\color[rgb]{0,0,0}$ab\omega_X d \epsilon_Y$}%
}}}}
\put(3511,1424){\makebox(0,0)[b]{\smash{{\SetFigFont{14}{16.8}{\familydefault}{\mddefault}{\updefault}{\color[rgb]{0,0,0}$a \neg c\omega_X$}%
}}}}
\put(3511,-1546){\makebox(0,0)[lb]{\smash{{\SetFigFont{14}{16.8}{\familydefault}{\mddefault}{\updefault}{\color[rgb]{0,0,0}same as above}%
}}}}
\put(4951,-151){\makebox(0,0)[lb]{\smash{{\SetFigFont{14}{16.8}{\familydefault}{\mddefault}{\updefault}{\color[rgb]{0,0,0}$ac \omega_X \omega_Y \neg d$}%
}}}}
\put(4951,839){\makebox(0,0)[lb]{\smash{{\SetFigFont{14}{16.8}{\familydefault}{\mddefault}{\updefault}{\color[rgb]{0,0,0}(one extension for each $\omega_Y$)}%
}}}}
\put(4951,1019){\makebox(0,0)[lb]{\smash{{\SetFigFont{14}{16.8}{\familydefault}{\mddefault}{\updefault}{\color[rgb]{0,0,0}$a \neg c \omega_X \omega_Y \neg d$}%
}}}}
\put(4951,1199){\makebox(0,0)[lb]{\smash{{\SetFigFont{14}{16.8}{\familydefault}{\mddefault}{\updefault}{\color[rgb]{0,0,0}OR}%
}}}}
\put(4951,1379){\makebox(0,0)[lb]{\smash{{\SetFigFont{14}{16.8}{\familydefault}{\mddefault}{\updefault}{\color[rgb]{0,0,0}$a \neg c\omega_X(d \vee \omega_Y)$}%
}}}}
\put(4951, 29){\makebox(0,0)[lb]{\smash{{\SetFigFont{14}{16.8}{\familydefault}{\mddefault}{\updefault}{\color[rgb]{0,0,0}OR}%
}}}}
\put(4951,209){\makebox(0,0)[lb]{\smash{{\SetFigFont{14}{16.8}{\familydefault}{\mddefault}{\updefault}{\color[rgb]{0,0,0}$ac\omega_X(d \vee \omega_Y)$}%
}}}}
\put(4951,-331){\makebox(0,0)[lb]{\smash{{\SetFigFont{14}{16.8}{\familydefault}{\mddefault}{\updefault}{\color[rgb]{0,0,0}(one extension for each $\omega_Y$)}%
}}}}
\end{picture}%
%%%%%%%%%%%%% end:cons-redu-3.latex %%%%%%%%%%%%%%
 %

\

The required defaults are the following ones. First, we
generate the considered model from the process that has
generated $ab \omega_X$:

\[
D_3 =
\left\{
\frac{b:\top}{d \wedge \neg y_1 \wedge \cdots \wedge \neg y_n}
\right\}
\]

From $a \neg c \omega_X$ and $ac\omega_X$ we generate an
arbitrary truth evaluation over $Y$. Since the model
$abd\omega_X\epsilon_Y$ assigns false to all variables
$y_i$, we cannot simply add $y_i$ as a conclusion. A similar
effect can be achieved by the following defaults.

\begin{eqnarray*}
D_4 &=&
\left\{
\left.
\frac{\neg c:\neg d \wedge y_i}{d \vee (y_i \wedge l_i)} ,~ 
\frac{\neg c:\neg d \wedge \neg y_i}{d \vee (\neg y_i \wedge l_i)}
\right|
1 \leq i \leq n
\right\}
\\
D_5 &=&
\left\{
\left.
\frac{c:\neg d \wedge y_i}{d \vee (y_i \wedge l_i)} ,~ 
\frac{c:\neg d \wedge \neg y_i}{d \vee (\neg y_i \wedge l_i)}
\right|
1 \leq i \leq n
\right\}
\end{eqnarray*}

The two defaults associated with $y_i$ and $\neg y_i$ cannot
be applied both at the same time, as the consequence of one
contains the negation of the justification of the other one.
Since the following defaults can only be applied when $d
\vee (l_1 \wedge \cdots \wedge l_n)$ has been derived, the
current extensions before their application are $a\neg c
\omega_X (d \vee (\omega_Y \wedge L))$ and $ac \omega_X (d
\vee (\omega_Y \wedge L))$, where $\omega_Y$ is an arbitrary
truth assignment over $Y$.

These extensions have all models of $ab \omega_X d
\epsilon_Y$. The following default removes these models from
the extensions if and only if $F$ is satisfiable for these
given assignments over $X$ and $Y$.

\[
D_6 =
\left\{
\frac{d \vee (l_1 \wedge \cdots \wedge l_n):\neg d \wedge F}{\neg d}
\right\}
\]

This default is not applicable from $ab\omega_Xd\epsilon_Y$
because its justification contains $\neg d$. It is
applicable from the other processes but only after the
$i$-th default of $D_4$ or $D_5$ has been applied for each
$i$ and only if the consequences of the applied defaults of
$D_4$ or $D_5$ are consistent with $\neg d \wedge F$. In
other words, $(d \vee (\omega_Y \wedge L)) \wedge \neg d
\wedge F$ must be consistent, which is equivalent to the
consistency of $\omega_Y \wedge F$ because $d$ and $L$ are
not mentioned in $\omega_Y$ and $F$.

We can therefore conclude that:

\begin{enumerate}

\item for each truth assignment $\omega_X$, three ``partial
extensions'' are generated from $W$: $a \neg c \omega_X$,
$ac \omega_X$, and $ab \omega_X$; the first two ones are
also generated by $W'$;

\item from $ab \omega_X$, the extension $ab \omega_X d
\epsilon_Y$ is generated; if the models of this extension
are not models of an extension of $W'$, equivalence between
$W$ and $W'$ does not hold;

\item from $a [\neg] c \omega_X$ we generate $a [\neg] c
\omega_X (d \vee (\omega_Y \wedge L))$ for each truth
evaluation $\omega_Y$ on the variables $Y$; to this formula,
$\neg d$ is added if and only if $F$ is consistent with
$\omega_X$ and $\omega_Y$.

\end{enumerate}

As a result, the models of $ab \omega_X d \epsilon_Y$ are
not models of an extension of $W'$ if and only if $F \wedge
\omega_X$ is satisfiable for every truth evaluation of $Y$.
Since non-equivalence has to be checked for every
$\omega_X$, we have that non-equivalence holds if and only
if $\exists X \forall Y \exists Z ~.~ F$.~\qed

A similar proof holds for constrained or rational default
logics by replacing the default theory of
Theorem~\ref{consequence-not-faithful} with that of
Theorem~\ref{consequence-not-faithful-global}. The proof can
also slightly simplified in this case, as the defaults of
$D_4$ and $D_5$ can be modified with justifications $y_i$ or
$\neg y_i$ and consequence $d \vee l_i$.

Since we have proved that the problem of clause redundancy
\wrt\  consequence-equivalence is both in \P{3}\  and hard
for the same class, we have the following theorem.

\begin{theorem}

The problem of checking whether $W' \equiv_D^c W$ is
\P{3}-complete if $W' \subseteq W$; hardness holds even if
$W = W' \cup \{a\}$.

\end{theorem}

%%%%%%%%%%%%% end:default-clause %%%%%%%%%%%%%%
 %

%%%%%%%%%%%%% begin:default-sets %%%%%%%%%%%%%%
\subsection{Complexity of Formula Redundancy}

The next problem to analyze is whether a formula (a set of
clauses) is redundant, for a fixed set of defaults. The
complexity of formula redundancy \wrt\  faithful and
consequence-equivalence is in \S{3}\ and \S{4},
respectively.

\begin{theorem}

The problem of formula redundancy for faithful and
consequence-equivalence is in \S{3}\  and \S{4},
respectively.

\end{theorem}

\proof Both problems can be expressed as the existence of a
subset $W' \subset W$ such that $W'$ is equivalent to $W$.
Since equivalence is in \P{2}\  and \P{3}, respectively, for
faithful and consequence-equivalence, the claim
follows.~\qed

Regarding hardness, we first show a theorem characterizing
the complexity of the problem for the case of faithful
equivalence. We then show a more general technique allowing
an hardness result to be raised one level in the polynomial
hierarchy.

\begin{theorem}

The problem of formula redundancy based on faithful
equivalence is \S{3}-hard.

\end{theorem}

\proof We reduce the problem of validity of $\exists X
\forall Y \exists Z . F$ to the problem of redundancy of a
formula. Let $n=|X|$. The default theory corresponding to
the formula $\exists X \forall Y \exists Z . F$ is the
theory $\l D,W \r$ defined as follows.

\begin{eqnarray*}
W &=&
\{s_i ~|~ 1 \leq i \leq n\} \cup
\{r_i ~|~ 1 \leq i \leq n\}
\\
\\
D &=& 
D_1 \cup D_2 \cup D_3 \cup D_4 \cup D_5 \cup D_6
\\
D_1 &=&
\left\{
\left.
\frac{s_i \wedge r_i:\neg s_j}{a} ,~
\frac{s_i \wedge r_i:\neg r_j}{a}
\right|
{1 \leq i \leq n \atop 1 \leq j \leq n}
\right\}
\\
D_2 &=&
\left\{
\left.
\frac{:\neg s_i \wedge \neg r_i}{a}
\right|
{1 \leq i \leq n \atop 1 \leq j \leq n}
\right\}
\\
D_3 &=&
\left\{
\left.
\frac{:y_i}{y_i\wedge h_i} ,~
\frac{:\neg y_i}{\neg y_i \wedge h_i}
\right|
1 \leq i \leq n
\right\} \\
\\
D_4 &=&
\left\{
\left.
\frac{:x_i}{p_i \wedge x_i} ~,~~
\frac{:\neg x_i}{p_i \wedge \neg x_i}
\right|
1 \leq i \leq n
\right\} \\
\\
D_5 &=&
\left\{
\left.
\frac{x_i \wedge r_i:\top}{\wedge W} ,~
\frac{\neg x_i \wedge s_i:\top}{\wedge W}
\right|
1 \leq i \leq n
\right\}
\\
D_6 &=&
\left\{
\frac{p_1 \ldots p_n h_1 \ldots h_n:F}
{\wedge W}
\right\}
\end{eqnarray*}

The defaults of $D_1$ and $D_2$ cannot be applied from $W$.
The defaults of $D_3$ and $D_4$ generates an extension for
every possible truth evaluation over $X \cup Y$; this
extension also contains all variables $h_i$ and $p_i$.
Whether or not the last default is applicable, its
consequence is equivalent to the background theory.

Let $W' \subset W$. If there is an index $i$ such that both
$s_i$ and $r_i$ are in $W'$, one of the defaults of $D_1$ is
applicable, generating $a$. Therefore, $W'$ is not
equivalent to $W$. If there exists an index $i$ such that
neither $s_i$ nor $r_i$ is in $W'$, the $i$-th default of
$D_2$ is applicable, still generating $a$.

In order to check for redundancy, we therefore only have to
consider subsets $W' \subset W$ for which either $s_i \in W$
or $r_i \in W$ but not both. Let $\omega_X$ be the assignment
on the variables $X$ such that $x_i$ is assigned to true if
$s_i \in W'$ and to false if $r_i \in W'$. The defaults of
$D_3$ and $D_4$ generate an arbitrary truth evaluation of
the variables $X \cup Y$. If the assignment on $X$ is not
equal to $\omega_X$, the formula $\wedge W$ is generated,
thus leading to an extension that is also an extension of
$W$. As a result, all extensions of $W'$ that do not match
the value $\omega_X$ are also extensions of $W$. If the same
holds also for the extensions for which the values of $X$
match $\omega_X$, then $W'$ is equivalent to $W$.

For a given $W'$ we consider the extensions consistent with
$\omega_X$. There is exactly one such extension for each
possible truth evaluation over $Y$. If the default of $D_6$
can be applied, it generates $\wedge W$, thus making $W'$
equivalent to $W$. In turn, the default of $D_6$ can be
applied for all truth evaluation over $Y$ if and only if for
all such truth evaluation, $F$ is satisfiable. As a result,
$W'$ is equivalent to $W$ if and only if, for all possible
truth evaluations over $Y$, the formula $F$ is satisfiable.
Since there exists a relevant $W'$ for each truth evaluation
over $X$, the formula $W$ is redundant if and only if there
exists a truth evaluation over $X$ such that, for all
possible truth evaluations over $Y$, the formula $F$ is
satisfiable.~\qed

%%%%%%%%%%%%% end:default-sets %%%%%%%%%%%%%%
 %

%%%%%%%%%%%%% begin:default-raise %%%%%%%%%%%%%%
\proofnewpage

In order to prove the \S{4}-hardness of the problem of
formula redundancy under consequence-equivalence, we should
provide a reduction from $\exists\forall\exists\forall$QBF
validity into this problem. A simpler proof can however be
given, based on the following consideration: checking clause
redundancy has been proved \P{2}-hard or \P{3}-hard using
reductions from QBFs that results in default theories having
$W=\{a\}$ as the background theory. As a result, these
reductions also prove that formula redundancy is \P{2}-hard
or \P{3}-hard. In other words, we can reduce the validity of
a $\forall\exists$QBF or a $\forall\exists\forall$QBF into
the problems of formula redundancy. What we show is that, if
such reductions satisfy some assumptions, we can obtain new
reductions from QBFs having an additional existential
quantifier in the front. The assumptions are that the
default theory resulting from the reduction is such that:

\begin{enumerate}

\item the background theory that results from the
reduction is classically irredundant;

\item the matrix of the QBF is only used in the
justification of a single default.

\end{enumerate}

The reductions used for proving the hardness of clause
redundancy satisfies both assumptions. In particular,
$\forall X \exists Y \forall Z . F$ is valid if and only if
the background theory of the following theory is
consequence-redundant, where $D$, $\alpha$, $\beta$,
$\gamma$, do not depend on $F$ but only on the quantifiers
of the QBF and $W$ is classically irredundant.

\[
\left\l
D \cup
\left\{
\frac{\alpha:\beta \wedge F}{\gamma}
\right\} ,~
W
\right\r
\]

The fact that the matrix of the QBF is copied ``verbatim''
in the default theory is exploited as follows: if $\omega_w$
is a truth evaluation over the variable $w$, then $\forall X
\exists Y \forall Z . F|_{\omega_w}$ is valid if and only if
the background theory of $\l D \cup \{ \frac{\alpha:\beta
\wedge F \wedge \omega_w}{\gamma} \} ,~ W \r$ is redundant.
This default theory can be modified in such a way the
subsets of the background theory are in correspondence with
the truth evaluations over $\omega_w$. This way, the
resulting theory is redundant if and only if $\exists w
\forall X \exists Y \forall Z ~.~ F$. The resulting default
theory still satisfies the two assumptions above on the
background theory and on the use of the matrix of the QBF;
therefore, this procedure can be iterated to obtain a
reduction from $\exists\forall\exists\forall$QBF validity
into the problem of formula redundancy under
consequence-equivalence. A similar technique can be used for
faithful equivalence.

\proofnewpage

The details of this technique are in the following three
lemmas. The first one shows that a literal can be moved from
the justification of a default to the background theory and
vice versa, under certain conditions.

\begin{lemma}
\label{just-to-back}

If the variable of the literal $l$ is not mentioned in $W$,
$D$, $\prec(d)$, and $\cons(d)$, the processes of the
following two theories are the same modulo the replacement
of $d$ with $d'$ and vice versa.

\begin{eqnarray*}
&& \l D \cup \{d\}, W \cup \{l\} \r \\
&& \l D \cup \{d'\}, W \r
\\
& \hbox to 0pt{where\hss} \\
&& d'=\frac{\prec(d):\just(d) \wedge l}{\cons(d)}
\end{eqnarray*}

\end{lemma}

\proof The literal $l$ and its negation only occur in the
background theory $W \cup \{l\}$ and in the justification of
$d$ and $d'$. The conditions on a process of the first
theory being selected either involve $(W \cup \{l\}) \cup
\just(d)$ or $W \cup \{l\}$ with other formulae not
containing $l$. As a result, moving $l$ from the background
theory to the justification of $d$ or vice versa does not
change these conditions.

Note that the processes are the same, but the extensions are
different in that $l$ is in all extensions of the first
theory but not in the extensions of the second.~\qed

The second lemma is an obvious consequence of the above:
under the same conditions, moving a literal from the
justification of a default to the background theory or vice
versa does not change the redundancy of a theory.

\begin{lemma}
\label{just-to-back-redu}

If $W' \subseteq W$, it holds $W' \equiv_{D'}^e W$ if and
only if $W' \cup \{l\} \equiv_{D''}^e W \cup \{l\}$, where
$l$ is a literal that is not mentioned in $W$, $D$,
$\prec(d)$, and $\cons(d)$, where $D'$ and $D''$ are as
follows.

\begin{eqnarray*}
D' &=&
D \cup
\left\{
\frac{\prec(d):\just(d) \wedge l}{\cons(d)}
\right\}
\\
D'' &=&
D \cup
\{d\}
\end{eqnarray*}

\end{lemma}

\proof Obvious consequence of the lemma above: $\l D', W'
\r$ and $\l D', W \r$ have the same processes of $\l D'', W'
\cup \{l\} \r$ and $\l D'', W \cup \{l\} \r$,
respectively.~\qed

A consequence of this lemma is that $W$ is redundant
according to $D'$ if and only if $W \cup \{l\}$ is redundant
according to $D''$. Indeed, $l$ is not mentioned in the
consequences of the defaults; therefore, a subset of $W \cup
\{l\}$ can only be equivalent to $W \cup \{l\}$ if it
contains $l$. The lemma is formulated in the more
complicated way because it is necessary for proving the
following lemma. The same property can be proved using
consequence-equivalence because moving $l$ from the
justification of the default to the background theory has
the only effect of adding $l$ to all extensions.

\proofnewpage

\begin{lemma}
\label{two-in-one}

If $W$ is classically irredundant,
then there exists $W' \subset W$ such that
$W' \cup \{w\} \equiv_D^e W \cup \{w\}$ or 
$W' \cup \{\neg w\} \equiv_D^e W \cup \{\neg w\}$
if and only if the following theory is redundant:

\begin{eqnarray*}
\left\l
D_w \cup
\left\{
\left.
\frac{p \wedge \alpha:\beta}{\gamma}
\right|
\frac{\alpha:\beta}{\gamma} \in D
\right\}
,~
W \cup \{w^+,w^-\}
\right\r
\end{eqnarray*}

\noindent where:

\begin{eqnarray*}
D_w =
\left\{
\frac{w^+ \wedge w^-:\neg W}{\neg p} ,~
\frac{:\neg w^+ \wedge \neg w^-}{\neg p} ,~
\frac{w^+:w \wedge p}{w \wedge p} ,~
\frac{w^-:\neg w \wedge p}{\neg w \wedge p}
\right\}
\end{eqnarray*}

\noindent and $w^+$, $w^-$, and $p$ are new variables.

\end{lemma}

\proof Since $w^+$ and $w^-$ are new variables not contained
in $W$ and $W$ is classically irredundant, $W \cup
\{w^+,w^-\}$ is classically irredundant as well.

We now consider the processes that can be generated from $W
\cup \{w^+, w^-\}$ and from its subsets. From $W \cup \{w^+,
w^-\}$ we can apply only one of the last two defaults of
$D_w$, generating either $w \wedge p$ or $\neg w \wedge p$.
From this point on, we have exactly the same processes of
$\l D,W \cup \{w\} \r$ and $\l D,W \cup \{\neg w\} \r$, the
generated extensions only differing because of the addition
of $p$ and $w^+$ or $w^-$.

The proper subsets of $W \cup \{w^+, w^-\}$ are $W' \cup
\{w^+, w^-\}$ where $W' \subset W$, $W' \cup \{w^+\}$, $W'
\cup \{w^-\}$, and $W'$, where $W' \subseteq W$. The fourth
subset $W'$ is not equivalent to $W$ because the second
default of $D_w$ allows the derivation of $\neg p$, which is
not derivable from $W$. If $W' \subset W$, since $W$ is
(classically) irredundant, $W' \cup \{w^+, w^-\}$ allows for
the application of the first default of $D_w$, deriving
$\neg p$; therefore, this subset is not equivalent to the
background theory.

The only two other subsets to consider are $W' \cup \{w^+\}$
and $W' \cup \{w^-\}$. In the first subset, only $w \wedge
p$ can be generated. In the second subset, only $\neg w
\wedge p$ can be generated. From this point on, we have
exactly the same processes of $W' \cup \{w\}$ and $W' \cup
\{\neg w\}$ according to $D$. The generated extensions are
the same but for the addition of $p$.~\qed

\proofnewpage

These three lemmas together proves that a reduction from QBF
to formula redundancy can be ``raised'' by the addition of
an existential quantifier in the front of the QBF.

\begin{lemma}
\label{raise}

If there exists a polynomial reduction from formulae $Q.E$,
where $Q$ is a sequence of quantifier of a given class, to
the problem of formula redundancy of a default theory in the
following form, then there exists a polynomial reduction
from formulae of the form $\exists w Q . F$ to the formula
redundancy of a theory in the following form.

\[
\left\l
D \cup
\left\{
\frac{\alpha:\beta \wedge F}{\gamma}
\right\}
,~
W
\right\r
\]

\noindent The formulae in $D$, $\{\alpha, \beta, \gamma\}$,
and $W$ do not depend on $F$. The background theory $W$ is
classically irredundant.

\end{lemma}

\proof Let $Q$ be a sequence of quantifiers so that the
validity of the formula $Q.E$ can be reduced to formula
redundancy of a default theory of the above form. We show a
reduction from the validity of $\exists w Q.F$ to formula
redundancy of a default theory of the same form.

By definition, both $Q.F|_{w=\true}$ and $Q.F|_{w=\false}$
can be reduced to the problem of formula redundancy. These
two formulae only differ on their matrixes, which are $
F|_{w=\true}$ and $F|_{w=\false}$. Therefore, the resulting
default theories are:

\begin{eqnarray*}
&&
\left\l
D \cup
\left\{
\frac{\alpha:\beta \wedge F|_{w=\true}}{\gamma}
\right\}
,~
W
\right\r
\\
&&
\left\l
D \cup
\left\{
\frac{\alpha:\beta \wedge F|_{w=\false}}{\gamma}
\right\}
,~
W
\right\r
\end{eqnarray*}

Since $w$ does not occur anywhere else in the theory, we can
replace $F|_{w=\true}$ and $F|_{w=\false}$ with $F \wedge w$
and $F \wedge \neg w$, respectively. Indeed, justifications
are only checked for consistency, and for any formula $R$
not containing $w$, the consistency of $R \cup F|_{w=\true}$
is the same as the consistency of $R \cup (F \wedge w)$, and
the consistency of $R \cup F|_{w=\false}$ is the same as the
consistency of $R \cup (F \wedge \neg w)$.

\begin{eqnarray*}
&&
\left\l
D \cup
\left\{
\frac{\alpha:\beta \wedge F \wedge w}{\gamma}
\right\}
,~
W
\right\r
\\
&&
\left\l
D \cup
\left\{
\frac{\alpha:\beta \wedge F \wedge \neg w}{\gamma}
\right\}
,~
W
\right\r
\end{eqnarray*}

By Lemma~\ref{just-to-back-redu}, formula redundancy of
these two theories corresponds to formula redundancy of the
same theories with $w$ or $\neg w$ moved to the background
theory. More precisely, the redundancy of the first theory
correspond to the existence of a subset $W' \subset W$ such
that $W' \cup \{w\}$ is equivalent to $W \cup \{w\}$
according to the defaults $D \cup \left\{\frac{\alpha:\beta
\wedge F}{\gamma}\right\}$. The same holds for the second
theory.

\begin{eqnarray*}
&&
\left\l
D \cup
\left\{
\frac{\alpha:\beta \wedge F}{\gamma}
\right\}
,~
W \cup \{w\}
\right\r
\\
&&
\left\l
D \cup
\left\{
\frac{\alpha:\beta \wedge F}{\gamma}
\right\}
,~
W \cup \{\neg w\}
\right\r
\end{eqnarray*}

By Lemma~\ref{two-in-one}, since $W$ is classically
irredundant, we have that either the first or the second of
the two theories are redundant if and only if the following
theory is redundant:

\[
\left\l
D_w \cup
\left\{
\left.
\frac{p \wedge \alpha:\beta}{\gamma}
\right|
\frac{\alpha:\beta}{\gamma} \in D
\right\}
,~
W \cup \{w^+,w^-\}
\right\r
\]

\noindent where $D_w$ is defined in the statement of
Lemma~\ref{two-in-one}. As a result, this formula is
redundant if and only if either $Q.F|_{w=\true}$ is valid or
$Q.F|_{w=\false}$ is valid, that is, $\exists w Q.F$ is
valid.

In order to complete the lemma, we have to show that the
background theory of the above theory is classically
irredundant, and the theory is in the form specified by the
statement of the theorem. Since $W$ is classically
irredundant by assumption and $w^+$ and $w^-$ are new
variables, $W \cup \{w^+,w^-\}$ is classically irredundant.
In the above theory, the matrix $F$ of the QBF is only
mentioned in the justification of the default $\frac{p
\wedge \alpha:\beta \wedge F}{\gamma}$. Therefore, the
theory that results from the reduction is in the form
specified by the theorem.~\qed

The above lemmas are also valid for consequence-equivalence.
In both cases, we have that the hardness of formula
redundancy is one level higher in the polynomial hierarchy
than clause redundancy.

\begin{theorem}

Formula redundancy is \S{3}-hard for faithful equivalence
and \S{4}-hard for consequence-equivalence.

\end{theorem}

\proof The reduction shown after Theorem~\ref{one-hard-exte}
and the reduction used in Theorem~\ref{one-hard-cons} are
reductions from $\forall\exists$QBF and
$\forall\exists\forall$QBF, respectively, into the problem
of formula redundancy. These reductions produce a default
theory in which the background theory contains a single
non-tautological clause, and is therefore irredundant, and
the matrix of the QBF only occurs in the justification of a
single default. These are the conditions of
Lemma~\ref{raise}. As a result, one can reduce an
$\exists\forall\exists$QBF or an
$\exists\forall\exists\forall$QBF to the problem of formula
redundancy by iteratively applying the modification of
Lemma~\ref{raise} for all variables of the first existential
quantifier.~\qed

\comment

should the hardness lemmas for clause redundancy state
that reduction satisfies these conditions?

\endcomment

%%%%%%%%%%%%% end:default-raise %%%%%%%%%%%%%%
 %

%%%%%%%%%%%%% begin:default-rule %%%%%%%%%%%%%%
\subsection{Redundancy of Defaults}

%%%%%%%%%%%%%%%%%%%%%%%%%%%%%%%%%%%%%%%%%%%%%%%%%%%%%%%%%%
% definitions

The redundancy of a default is defined in the same way as
redundancy of clauses.

\begin{definition}
[Redundancy of a default]

A default $d$ is redundant in $\l D,W \r$
if and only if 
$\l D \backslash \{d\},W \r$ is equivalent to $\l D,W \r$.

\end{definition}

This definition depends on the kind of equivalence used.
Therefore, a default can be redundant \wrt\  faithful or
consequence-equivalence. The redundancy of defaults is
defined as follows.

\begin{definition}
[Redundancy of a theory]

A default theory $\l D,W \r$ is default redundant if and
only if there exists $D' \subset D$ such that $\l D',W \r$
is equivalent to $\l D,W \r$.

\end{definition}

%%%%%%%%%%%%%%%%%%%%%%%%%%%%%%%%%%%%%%%%%%%%%%%%%%%%%%%%%%
% making defaults irredundant: necessary even for proving
% that local redundancy is not the same as global

\subsubsection{Making Defaults Irredundant}

The following lemma is the version of
Theorem~\ref{default-irredundant} to the case of default
redundancy rather than clause redundancy. It proves that
some defaults can be made irredundant while not changing the
redundancy status of the other ones.

\begin{lemma}
\label{default-default-irredundant}

For every default theory $\l D,W \r$, set of defaults $D_I
\subseteq D$, and $D_1$, $D_2$, $D_3$ defined as follows:

\begin{eqnarray*}
D_1 &=& \{d+, d-\} \\
& \hbox to 0pt{where:\hss} \\
&& d+ = \frac{:p \wedge q}{p \wedge q} \\
&& d- = \frac{:\neg p \wedge q}{\neg p \wedge q} \\
D_2 &=&
\left\{
\left.
\frac{q \wedge (\neg p \vee \alpha): \neg p \vee \beta}
{(p \vee v_i) \wedge (\neg p \vee \gamma)} 
\right|
\frac{\alpha:\beta}{\gamma} \in D_I
\right\} \\
D_3 &=&
\left\{
\left.
\frac{q \wedge p \wedge \alpha:\beta}{\gamma}
\right|
\frac{\alpha:\beta}{\gamma} \in D \backslash D_I
\right\}
\end{eqnarray*}

\noindent if $\l D,W \r$ has extensions and $W$ is
consistent, it holds that:

\begin{enumerate}

\item the processes of $\l D_1 \cup D_2 \cup D_3,W \r$ are
(modulo the transformation of the defaults) the same of $\l
D,W \r$ with $d+$ added to the front and a number of
processes composed of $d-$ and a sequence containing all
defaults of $D_2$;

\item the extension of $\l D_1 \cup D_2 \cup D_3,W \r$ are
the same of $\l D,W \r$ with $\{p,q\}$ added plus the single
extension $\{\neg p,q\} \cup \{v_i\}$;

\item a subset of $D_1 \cup D_2 \cup D_3$ is equivalent to
it if and only if it contains $D_1 \cup D_2$ and the set of
original defaults corresponding to those of $D_2 \cup D_3$ is
equivalent to $D$.

\end{enumerate}

\end{lemma}

\proof Since all defaults of $D_2 \cup D_3$ have $q$ as a
precondition, they are not applicable from $W$. The only
defaults that are applicable to $W$ are therefore $d+$ and
$d-$, which are mutually exclusive.

Let us consider the processes with $d-$ in first position.
Since $d-$ generates $\neg p$, the defaults of $D_3$ are not
applicable. We prove that $[d-] \cdot \Pi_2$ is a successful
process, where $\Pi_2$ is an arbitrary sequence containing
all defaults of $D_2$. The preconditions of all defaults of
$D_2$ are entailed by $q \wedge \neg p$. The union of the
justifications and consequences of all defaults of this
process is $\{\neg p, q\} \cup \{\neg p \vee \beta, p \vee
v_i, \neg p \vee \gamma\}$, which is equivalent to $\{\neg
p, q\} \cup \{v_i\}$. This set is consistent with the
background theory, which does not contain the variables $p$,
$q$, and $v_i$.

If a subset of $D_1 \cup D_2 \cup D_3$ does not contain
$d-$, the literal $\neg q$ cannot derived because no other
default has $\neg q$ as a conclusion. If a subset of $D_1
\cup D_2 \cup D_3$ does not contain a default of $D_2$, the
corresponding variable $v_i$ is not in this extension. As a
result, every subset of $D_1 \cup D_2 \cup D_3$ that is
equivalent to it contains $\{d-\} \cup D_2$.

Let us now consider the processes with $d+$ in first
position. Such a process cannot contain $d-$. Since $p$ and
$q$ are generated, the defaults of $D_2 \cup D_3$ can be
simplified to $\frac{\alpha:\beta}{\gamma}$ by removing all
clauses containing $p$ or $q$ and all literals $\neg p$ and
$\neg q$ from the clauses containing them. As a result, the
processes having $d+$ in first position correspond to the
processes of the original theory.

Provided that the original theory has extensions, every
subset of $D_1 \cup D_2 \cup D_3$ not containing $d+$ lacks
these extensions. The defaults of $D_3$ are redundant if and
only if they are redundant in the original theory. More
precisely, a subset $D' \subset D_1 \cup D_2 \cup D_3$ is
equivalent to $D_1 \cup D_2 \cup D_3$ if and only if $D'$
contains $D_1 \cup D_2$, and the set of original defaults
$D''$ corresponding to the defaults of $D' \cap (D_2 \cup
D_3)$ is equivalent to $D$.~\qed

%%%%%%%%%%%%%%%%%%%%%%%%%%%%%%%%%%%%%%%%%%%%%%%%%%%%%%%%%%
% global redundancy != local redundancy

\subsubsection{Redundancy of Defaults vs.\ Sets of Defaults}

While a formula is classically redundant if and only if it
contains a redundant clause, the same does not happen for
default redundancy. The following theorem indeed proves that
Reiter and rational default logic do not have the local
redundancy property \wrt\  redundancy of defaults.

\begin{theorem}

There exists a set of defaults $D$ such that, according to
Reiter and rational default logic:

\begin{enumerate}

\item for any $d \in D$, the theory $\l D \backslash
\{d\},\emptyset \r$ has extensions and $\l D \backslash
\{d\},W \r \not\equiv_D^c \l D,\emptyset \r$;

\item there exists $D' \subset D$ such that
$\l D',\emptyset \r \equiv_D^e \l D,\emptyset \r$.

\end{enumerate}

\end{theorem}

\proof We use a pair of defaults that lead to failure is
they are together in the same process. Removing one of them
from the default theory leads to a new extension, while
removing both of them lead to the original set of
extensions. The following defaults are a realization of this
idea.

\begin{eqnarray*}
D &=& \{d_1,d_2,d_3\} \\
& \hbox to 0pt{where:\hss} \\
&& d_1 = \frac{:b}{b \wedge c} \\
&& d_2 = \frac{:b}{b \wedge \neg c} \\
&& d_3 = \frac{:\neg b}{\neg b}
\end{eqnarray*}

The extensions of some $\l D',\emptyset \r$, with $D'
\subseteq D$, are as follows:

\begin{description}

\item[$D'=D$] we can either apply both $d_1$ and $d_2$
(leading to a failure) or $d_3$ alone; the only extension of
this theory is therefore $\neg b$;

\item[$D'=\{d_1,d_3\}$] both $d_1$ and $d_3$ can be applied,
but not both; that results in two processes having
conclusions $\vee \ext(\l D,\emptyset \r)= (b \wedge \neg c)
\vee \neg b \equiv \top$;

\item[$D'=\{d_2,d_3\}$] same as above: $\vee \ext(\l
D,\emptyset \r)= (b \wedge c) \vee \neg b \equiv \top$;

\item[$D'=\{d_3\}$] the only selected process is $[d_3]$,
which leads to $\vee \ext(\l D',W \r)=\neg b$.

\end{description}

As a result, $\l D \backslash \{d\},\emptyset \r$ has
extensions for every $d \in D$. The default $d_3$ is not
irredundant, but can be made so by the transformation of
Theorem~\ref{default-default-irredundant}, which preserves
processes almost exactly; an alternative is to replace $d_3$
with $\frac{:\neg b}{b \wedge d}$ and $\frac{:\neg b}{b
\wedge e}$. The resulting set of default has no redundant
default, but has an equivalent subset.~\qed

The same result holds for constrained default logic.

\begin{theorem}

There exists a set of defaults $D$ such that, according to
constrained default logic:

\begin{enumerate}

\item for any $d \in D$, it holds $\l D \backslash \{d\},W
\r \not\equiv_D^c \l D,\emptyset \r$;

\item there exists $D' \subset D$ such that
$\l D',\emptyset \r \equiv_D^e \l D,\emptyset \r$.

\end{enumerate}

\end{theorem}

\proof The defaults are the following ones:

\begin{eqnarray*}
D &=& \{d_1,d_2,d_3\} \\
& \hbox to 0pt{where:\hss}\\
&& d_1 = \frac{:x}{a} \\
&& d_2 = \frac{:x}{b} \\
&& d_3 = \frac{:\neg x \wedge \neg y}{ab}
\end{eqnarray*}

The theory $\l D,\emptyset \r$ has two selected processes
(modulo permutation of defaults): $[d_1,d_2]$ and $[d_3]$,
both generating the extension $ab$. Removing either $d_1$ or
$d_2$ causes the first process to become $[d_1]$ or $[d_2]$,
thus creating a new extension that is either $a$ or $b$. On
the other hand, removing both $d_1$ and $d_2$ makes the only
remaining process to be $[d_3]$, which generates the only
extension $ab$ of the original theory. The default $d_3$ is
not redundant, but can be made so by applying the
transformation of
Theorem~\ref{default-default-irredundant}.~\qed

\proofnewpage

Justified default logic has the local redundancy property
\wrt\  default redundancy. This is a combination of two factors:
first, justified default logic is failsafe
\cite{libe-failsafe} (every successful process can be made
selected by adding some defaults); second, every extension
is generated by an unique set of defaults. The proofs
requires two lemmas. The first one is about extendibility of
processes when new defaults are added to a theory.

\begin{lemma}
\label{justified-monotonic}

In justified default logic, if $\Pi$ is a selected process
of $\l D',W \r$ and $D' \subseteq D$, then there exists a
sequence $\Pi'$ of defaults of $D \backslash D'$ such that
$\Pi \cdot \Pi'$ is a selected process of $\l D,W \r$.

\end{lemma}

\proof Let $\Pi$ be a selected process of $\l D',W \r$. By
definition, it holds $W \cup \cons(\Pi[d]) \models \prec(d)$
and $W \cup \cons(\Pi) \top \just(d)$ for every $d \in \Pi$.
As a result, $\Pi$ is a also a successful process of $\l D,W
\r$. Therefore, there exists $\Pi'$ such that $\Pi \cdot
\Pi'$ is a selected process of $\l D,W \r$ because justified
default logic is failsafe \cite{libe-failsafe}. If $\Pi'$
contains defaults of $D'$, then $\Pi$ would not be a closed
process of $\l D',W \r$.~\qed

In order for proving the second lemma, we need an
intermediate result, which is already well known.

\begin{lemma}

In justified default logic, the selected processes of $\l
D,W \r$ generating the extension $E$ are composed of exactly
the defaults of the following set:

\[
GEN(E,D) =
\{d \in D ~|~ E \models \prec(d) \mbox{ and }
E \top \just(d) \cup \cons(d) \}
\]

\end{lemma}

\proof Assume that $\Pi$ is a selected process generating
$E$ that does not contain a default $d \in GEN(E,D)$. Since
$E \models \prec(d)$, $E \top \just(d) \cup \cons(d)$, and
$E = W \cup \cons(\Pi)$, we have that $W \cup \cons(\Pi)
\models \prec(d)$ and $W \cup \cons(\Pi) \top \just(d) \cup
\cons(d)$. As a result, $\Pi \cdot [d]$ is a successful
process, contradicting the assumption.

Let $\Pi$ be a selected process containing a default $d$ not
in $GEN(E,D)$. By definition, either $E \not\models
\prec(d)$ or $E \bot \just(d) \cup \cons(d)$. The first
condition implies that $W \cup \cons(\Pi[d]) \not\models d$
whichever the position of $d$ in $\Pi$ is. The second
condition implies $W \cup \cons(\Pi) \bot \just(d) \cup
\cons(d)$: the process $\Pi$ is not successful contrary to
the assumption.~\qed

The next lemma relates the processes of two theories when
they are assumed to have the same extension. In this lemma
and in the following theorem, when a process is used in a
place where a set of defaults is expected, it means the set
of defaults of the process. For example, if $\Pi$ is a
sequence of defaults and $D'$ a set of defaults, $\Pi \cap
D'$ is the set of defaults that are both in $\Pi$ and in
$D$.

\begin{lemma}
\label{justified-subset}

In justified default logic, if $D' \subseteq D$, $\l D',W \r
\equiv_D^e \l D,W \r$, and $\Pi$ is a selected process $\l
D,W \r$, then there exists a selected process of $\l D',W
\r$ made exactly of the defaults of $\Pi \cap D'$ and
generating the same extension generated by $\Pi$.

\end{lemma}

\proof Let $E = W \cup \cons(\Pi)$ be the extension that is
generated by $\Pi$. By the lemma above, it is generated by
the defaults in $GEN(E,D)$. Since $E$ is also an extension
of $\l D',W \r$, it is generated by a process $\Pi'$ made
exactly of the defaults of $GEN(E,D') = GEN(E,D) \cap E' =
\Pi \cap D'$.~\qed

\proofnewpage

The main theorem relating the extensions of theories
differing for the set of defaults in justified default logic
is the following one.

\begin{theorem}

If $D' \subseteq D'' \subseteq D$ and
$\l D',W \r \equiv^e \l D,W \r$ then
$\l D',W \r \equiv^e \l D'',W \r$ for justified default logic.

\end{theorem}

\proof We first show that every extension of $\l D'',W \r$
is also an extension of $\l D,W \r$, and then show the
converse.

Let $E$ be an extension of $\l D'',W \r$. Let $\Pi$ is one
its generating processes. By
Lemma~\ref{justified-monotonic}, there exists a sequence
$\Pi'$ of defaults of $D \backslash D''$ such that $\Pi
\cdot \Pi'$ is a selected process of $\l D,W \r$. Let $E'$
be its generated extension. Since $E$ is generated by $\Pi$
and $E'$ is generated by $\Pi \cdot \Pi'$, we have $E'
\models E$. We prove that $E \models E'$, which implies $E
\equiv E'$.

By Lemma~\ref{justified-subset}, since $\Pi$ is a selected
process of $\l D,W \r$ and this theory is faithfully
equivalent to $\l D',W \r$, there exists a selected process
$\Pi''$ of $\l D',W \r$ made of the defaults of $(\Pi \cdot
\Pi') \cap D'$ and generating the extension $E'$. Since
$\Pi'$ is made of defaults of $D \backslash D''$ and $D'
\subseteq D''$, we have that $(\Pi \cdot \Pi') \cap D' = \Pi
\cap D'$. As a result, $\Pi''$ is only made of defaults in
$\Pi \cap D'$. Since $\Pi''$ generates $E'$ and $\Pi$
generates $E$, we have $E \models E'$. We can therefore
conclude that $E \equiv E'$.

\

Let us now prove the converse: we assume that $E$ is an
extension of $\l D,W \r$ and prove that it is also an
extension of $\l D'',W \r$. Let $\Pi$ be the process of $\l
D,W \r$ that generates $E$. By definition, the following two
properties are true:

\begin{enumerate}

\item $W \cup \cons(\Pi) \models \prec(d)$ for every $d \in
\Pi$;

\item $W \cup \cons(\Pi) \bot \just(d) \cup \cons(d)$ for
every $d \not \in \Pi$.

\end{enumerate}

By Lemma~\ref{justified-subset}, the theory $\l D',W \r$ has
a selected process $\Pi'$ that is composed exactly of the
defaults of $\Pi \cap D'$ and that generates the same
extension $E$. Since $W \cup \cons(\Pi') \equiv W \cup
\cons(\Pi)$, the two properties are equivalent to the
following two ones:

\begin{enumerate}

\item $W \cup \cons(\Pi') \models \prec(d)$ for every $d \in
\Pi$;

\item $W \cup \cons(\Pi') \bot \just(d) \cup \cons(d)$ for
every $d \not \in \Pi$.

\end{enumerate}

The first property implies that every default $d \in \Pi
\cap (D'' \backslash D')$ is applicable to $\Pi'$: this is
because the precondition of $d$ is entailed by $W \cup
\cons(\Pi')$ and the process $\Pi' \cdot [d]$ is successful
because so is $\Pi$, which contains all default of $\Pi'
\cdot [d]$. The second property implies that no default of
$D'' \backslash \Pi$ is applicable to $\Pi'$. As a result,
$\Pi'$ and the sequence composed of all defaults of $\Pi
\cap (D'' \backslash D')$ in any order form a selected
process of $D''$. The extension generated by this process is
equivalent to $E$ because this process is composed of a
superset of the defaults of $\Pi' $ and a subset of the
defaults of $\Pi$, and these two processes both generate
$E$.~\qed

We therefore have as a corollary that justified default
logic has the local redundancy property when redundancy of
defaults is considered.

\begin{corollary}

Justified default logic has the local redundancy property
\wrt\  redundancy of defaults.

\end{corollary}

%%%%%%%%%%%%%%%%%%%%%%%%%%%%%%%%%%%%%%%%%%%%%%%%%%%%%%%%%%
% redundancy of clauses -> defaults

\subsubsection{Redundancy of Clauses and of Defaults}

For Reiter and rational default logic, an upper bound on
complexity can be given by showing a reduction from the
complexity of clause or formula redundancy to the
corresponding problems for defaults. This is possible thanks
to the following lemma.

\begin{lemma}
\label{clause-default}

$\l D,W \cup \{\gamma\} \r$ has the same Reiter and rational
extensions of $\l D \cup \{d_\gamma\},W \r$, where
$d_\gamma=\frac{:\top}{\gamma}$.

\end{lemma}

\proof Since $d_\gamma$ has no precondition and a
tautological justification, it is always applicable.
Therefore, every process of $\l D \cup \{d_\gamma\},W \r$
contains this default, and therefore generates
$\gamma$.~\qed

This lemma can be iterated for all clauses of $W$, leading
to the following result.

\begin{theorem}

For Reiter and rational default logic, the problems of
checking the redundancy of a default or of a default theory
are at least as hard as the corresponding problems for clause
redundancy.

\end{theorem}

\proof The clause $\gamma$ is redundant in $\l D,W \r$ if
and only if $d_\gamma$ is redundant in $\l D \cup
\{d_\gamma\},W \backslash \{\gamma\} \r$. Indeed, $\l D,W
\r$ has the same extensions of $\l D \cup \{d_\gamma\},W
\backslash \{\gamma\} \r$, and removing $\gamma$ from the
first theory or removing $d_\gamma$ from the second theory
lead both to $\l D,W \backslash \{\gamma\} \r$.

The problems of formula redundancy can be reduced to default
redundancy by first applying Lemma~\ref{clause-default} to
all clauses of $W$, and then making all original defaults
irredundant using the transformation of
Lemma~\ref{default-default-irredundant}.~\qed

The complexity of redundancy for defaults can be therefore
characterized as follows.

\begin{corollary}

For Reiter and justified default logic, the problem of
redundancy of a default is \P{2}-hard and \P{3}-hard for
faithful and consequence-equivalence, respectively; the
problem of redundancy of a default theory is \S{3}-hard and
\S{4}-hard for faithful and consequence-equivalence,
respectively.

\end{corollary}

Equivalence of extensions can be proved to be in \P{2}\
even if the defaults or the background theories are not even
related.

\begin{theorem}

Checking whether $\l D,W \r \equiv_D^e \l D',W' \r$
is in \P{2}\  for Reiter and justified default logic.

\end{theorem}

\proof The contrary of the statement amounts to checking
whether any of the two theories have an extension that the
other one does not have. The number of possible extensions,
however, is limited by the fact that any extension is
generated by the set of consequences of some defaults.

Checking whether $\l D,W \r$ has an extension that $\l D',W'
\r$ has not can be done as follows: guess a subset $D''
\subseteq D$, and let $E=\cons(D'')$; check whether $E$ is
an extension of $\l D,W \r$ but is not an extension of $\l
D',W' \r$.

Checking whether a formula $E$ is an extension of a default
theory can be done with a logarithmic number of
satisfiability tests \cite{rosa-99,libe-exte}. As a result,
the problem can also be expressed as a QBF formula
$\exists\forall$QBF. In order to check whether there exists
$D''$ such that $E = \cons(D'')$ is in this condition, we
only have to add an existential quantifier to the front of
this formula. The problem is therefore in \P{2}.~\qed

The problem of checking the default redundancy of a theory
is obviously in \S{3}, as it can be solved by guessing a
subsets of defaults and then checking equivalence.

\begin{corollary}

The problem of checking the redundancy of a default or the
default redundancy of a theory are \P{2}-complete and
\S{3}-complete, respectively, for Reiter and justified
default logic for faithful equivalence.

\end{corollary}

Consequence equivalence can also be proved to have the same
complexity as for the case studied for clauses.

\begin{theorem}

Checking the consequence-equivalence for Reiter and
justified default logic is in \P{3}.

\end{theorem}

\proof The converse of the problem can be expressed as:
there exists a model $M$ that is a model of an extension of
the first theory but not of the second, or vice versa. This
corresponds to two quantifications over extensions and a
check for whether a formula is an extension. The latter is
in \Dlog{2}\  for the two considered semantics
\cite{rosa-99,libe-exte}. Therefore, the whole problem is in
\P{3}.~\qed

As a consequence, the complexity of redundancy for
consequence-equivalence is exactly characterized for Reiter
and justified default logics.

\begin{corollary}

The problem of checking the redundancy of a default or the
default redundancy of a theory are \P{3}-complete and
\S{4}-complete, respectively, for Reiter default logic and
consequence-equivalence.

\end{corollary}

\comment

open problem:

reduction for failsafe semantics

upper bound for constrained and rational

\endcomment

%%%%%%%%%%%%% end:default-rule %%%%%%%%%%%%%%
 %

%%%%%%%%%%%%% end:default %%%%%%%%%%%%%%
 %

\typeout{}
\typeout{Acks!}
\typeout{}
%%%%%%%%%%%%% begin:acks %%%%%%%%%%%%%%
\subsection*{Acknowledgments}

The author thanks Jiang Zhengjun for his comments about on
this paper.

%%%%%%%%%%%%% end:acks %%%%%%%%%%%%%%
 %

\bibliographystyle{alpha}

\end{document}